\documentclass[useAMS,usenatbib]{mn2e}

\usepackage{graphicx}
\usepackage{times}
\usepackage{amssymb,amsmath}
\usepackage{hyperref}

\newcommand{\beq}{\begin{equation}}
\newcommand{\eeq}{\end{equation}}

\newcommand{\lapprox}{$\stackrel {<}{_{\sim}}$}
\newcommand{\lsim}{\ \raise
-2.truept\hbox{\rlap{\hbox{$\sim$}}\raise5.truept\hbox{$<$}\ }}
\newcommand{\gsim}{\ \raise
-2.truept\hbox{\rlap{\hbox{$\sim$}}\raise5.truept\hbox{$>$}\ }}
\newcommand{\simsim}{\ \raise
-2.truept\hbox{\rlap{\hbox{$\sim$}}\raise5.truept\hbox{$\sim$}\ }}

\def\gtorder{\mathrel{\raise.3ex\hbox{$>$}\mkern-14mu
                \lower0.6ex\hbox{$\sim$}}}
\def\ltorder{\mathrel{\raise.3ex\hbox{$<$}\mkern-14mu
                \lower0.6ex\hbox{$\sim$}}}

\def\arcmin{\hbox{$^\prime$}}
\def\arcsec{\hbox{$^{\prime\prime}$}}
\def\solar{\mbox{$_{\normalsize\odot}$}}

\def\aj{AJ}                   
\def\araa{ARA\&A}             
\def\apj{ApJ}                 
\def\apjl{ApJL}                
\def\apjs{ApJS}

\def\aap{A\&A}

\def\mnras{MNRAS}

\def\pasp{PASP}

\def\ssr{Space~Sci.~Rev.}

\def\nar{New~Astron.~Rev.}

\title[Bimodal Stellar Clustering in NGC\,346]{The Complex Distribution of Recently Formed Stars.\\ Bimodal Stellar Clustering in the Star-Forming Region NGC\,346.} 

\author[D. A. Gouliermis et al.]
  {Dimitrios A.~Gouliermis,\thanks{dgoulier@mpia.de; dgoulierm@googlemail.com}
  Sacha Hony, and 
 Ralf S. Klessen
  \\
 Universit\"at Heidelberg, Zentrum f\"ur Astronomie, Institut f\"ur Theoretische Astrophysik, 
 Albert-Ueberle-Str.~2, 69120 Heidelberg, Germany\\
  }
  
\begin{document}

\date{To be submitted to MNRAS}

\pagerange{\pageref{firstpage}--\pageref{lastpage}} \pubyear{2013}

\maketitle

\label{firstpage}


\begin{abstract}
We present a detailed stellar clustering analysis with the application of the two-point correlation function on distinct young stellar ensembles. 
Our aim is to understand how stellar systems are assembled at the earliest stages of their formation. Our object of interest is the star-forming region NGC\,346 in the 
Small Magellanic Cloud. It is a young stellar system well-revealed from its natal environment, comprising complete samples of pre--main-sequence and upper main-sequence stars, very close to their formation. We apply a comprehensive characterization of the autocorrelation function for both centrally condensed stellar clusters and  self-similar stellar distributions through numerical simulations of stellar ensembles. We interpret the observed autocorrelation function of 
NGC\,346 on the basis of these simulations. We find that it can be best explained as the combination of two distinct stellar clustering designs, a centrally concentrated, dominant at the central part of the star-forming region, and an extended self-similar distribution of stars across the complete observed field. The cluster component, similar to non-truncated young star clusters, is determined to have a core radius of $\sim$\,2.5\,pc and a density profile index of $\sim $\,2.3. The extended fractal component is found with our simulations to have a fractal dimension of $\sim$\,2.3, identical to that found for the interstellar medium, in agreement to hierarchy induced by turbulence. This suggests that the stellar clustering at a time very near to birth behaves in a complex manner. It is the combined result of the star formation process regulated by turbulence and the early dynamical evolution induced by the gravitational potential of condensed stellar clusters. 
\end{abstract}




\section{Introduction}

It is generally accepted that stars form in
groups of various sizes and characteristics \citep{ladalada03},
starting with small compact concentrations of protostars embedded in
star-forming regions and moving up in {\sl length-scale} to large
extended loose aggregates of young stars
stars. It is, however,
suggested that these diverse stellar assembles are not independent
from each other, but tightly connected through the star formation
process \citep{elmegreen11}. Small, dense proto-clusters coexist in a
symbiotic fashion with larger, less dense subgroups of OB-type stars,
which in turn reside in even larger, looser stellar associations, each
of these types of objects representing a different {\sl time-scale} of star 
formation within one molecular cloud \citep{efremovelmegreen98}.
This picture of clustered star formation at various length- and time-scales 
is not always clear in our observations, as e.g., in the massive star formation
environments of star-burst clusters. 

Studies of newly formed stellar systems can identify the conditions that may 
favor {\sl multiple} over {\sl single} cluster formation events.
Different scenarios for star formation predict different
observable properties for the resulting stellar systems. 
The `quiescent' star formation scenario \citep[e.g.][]{krumholztan07}
predicts large age-spreads among young stars in the same molecular
cloud, while according to the `competitive accretion' scenario
\citep[e.g.][]{clark07} star formation is a process of short
time-scale, leading to clusters in a variety of forms.  

The investigation of the clustering of stars at the time of their
formation can provide important information on the nature of the star
formation process itself \citep[e.g.,][]{schmeja08}, and place constraints 
to the suggested theories. A single compact
stellar cluster, embedded in its own H{\sc ii} region, would suggest a
local monolithic episode of star formation in a dense environment. 
However, such clusters are rarely found in isolation; they are the densest 
stellar concentrations of larger stellar structures as seen in dwarf and spiral 
galaxies \citep{efremov09, karampelas09}. In our own Galaxy multiple (or
fractured) clusterings of stars in large star-forming regions, are
also more commonly observed \citep{feigelson11, megeath12}. This
clustering behavior favors the existence of multiple processes, such
as feedback, controlling star formation, and being active on different
scales.


Giant Molecular Clouds are hierarchical structures \citep{elmegreen96, stutzki98},
 indicating that scale-free processes  determine their global morphology.
Turbulence is being widely accepted as the dominant among these
processes \citep{maclowklessen04, elmegreenscalo04}. The goal of this paper is to establish 
whether the newly formed stars follow a similar, hierarchical distribution, 
which may indicate that turbulence also determines the clustering behavior 
of stars at the time of their formation. This investigation requires a complete 
census of stars, covering a high dynamic range in masses, distributed over the 
typical length-scale of giant molecular clouds. 


Our target of interest is the star-forming complex NGC\,346, the brightest {\sc H\,ii} region in the Small Magellanic Cloud 
\citep[LHA\,115-N66;][]{henize56}. {\sl Hubble} Space Telescope imaging of such regions in the Magellanic Clouds (MCs) 
provides an unprecedented access to their newly-born stellar populations down to the sub-solar regime 
over large areas of the sky. Observed by {\sl Hubble} (GO Program 10248; PI: A. Nota), this region satisfies, thus, the main observational
criteria for our analysis: 1) Observed on size-scales relevant for molecular clouds (50\,-\,100\,pc); 2) High angular resolution 
($\sim$\,0.125\arcsec); 3) Large number of detected members, covering a significant fraction of the stellar Initial Mass Function,
complete to $\sim$\,0.5 M{\solar}. 
Our dataset, obtained with the Advanced Camera for Surveys (ACS), is described in \cite{gouliermis06}. 

\begin{figure}
\centering
 \includegraphics[width=\columnwidth]{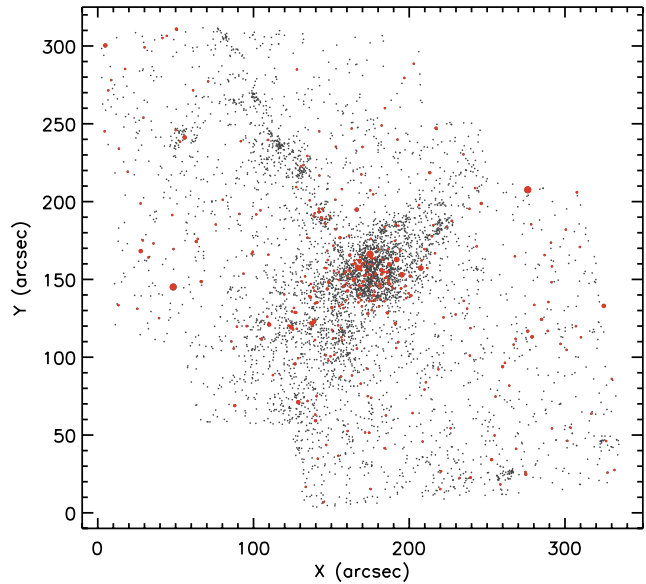} 
\caption{Map of the stars selected for our analysis, i.e., low-mass PMS and more massive upper--main-sequence 
(UMS) stars, as found in both F555W and F814W filters with {\sc dolphot} photometry \citep{dolphin00} based 
on ACS/WFC imaging of three pointings on the field of NGC\,346. Red dots correspond to the UMS stars. 
The map covers the whole observed area. North is up, and east is to the left. Coordinates are given in seconds of arc in respect to
a reference point, and are determined based on the drizzled ACS image in the F814W filter.}
\label{f:map}
\end{figure}

A rich sample of more than 98,000 stars was detected down to $m_{\rm 555} \simeq 27$~mag with \gsim 
50\%  completeness. It comprises a mixture of stellar generations, with 60\% of the stars formed 
\lsim\,5\,Gyr ago \citep{cignoni11}. The young populations in NGC\,346 consist mainly of low-mass pre--main-sequence 
(PMS) stars, identified from their positions on the color-magnitude diagram  \citep[see][for a review on low-mass PMS stars in the Magellanic Clouds] {gouliermis12}. It is decidedly troublesome 
to determine an age for a cluster based on its PMS stars \citep[][]{jeffries12, preibisch12}. 
Nevertheless, an isochronal age of $\sim$\,3\,Myr has been established for NGC\,346 by \cite{sabbi08} by 
comparison with PMS evolutionary models. We complete the sample of low-mass PMS stars with the upper--main-sequence 
(UMS) stars ($m_{\rm 555}-m_{\rm 814} \leq$\,0.0\,mag; 12\lsim\,$m_{\rm 555}$\lsim\,17\,mag; ages \lsim\,10\,Myr), 
compiling a total sample of 5,150 stars. The UMS stars correspond to about 7\% of the sample. The  map of our stellar 
inventory  is shown in Figure\,\ref{f:map}.

In a previous study \citep[][from here on Paper\,I]{schmeja09} we
applied a cluster analysis based on the nearest-neighbor density method
and we identified ten individual PMS stellar clusters in the region. We established
 that NGC\,346 is a multi-clustered environment. 
This work also provided evidence of hierarchy in the PMS stellar
clustering from a graph theory study with the minimum spanning tree 
and the analysis of the $\cal{Q}$ parameter \citep{cw04}.
While these methods provided a unprecedented insight of the NGC\,346 clustering,
they were not able to quantitatively describe its complexity, or to characterize its 
self-similar behavior. In addition, the accuracy of the $\cal{Q}$ parameter in 
interpreting a fractal structure has been challenged on the basis of the effect of 
projection in elongated star clusters (see \citealt{bastian09} and \citealt{cw09} for 
different accessions and viewpoints to the problem). The $\cal{Q}$ parameter 
for the whole complex of NGC\,346 was found equal to about 0.8, and thus cannot 
be used to conclude about the nature of stellar clustering in this region.


 We revisit the question of stellar clustering in NGC\,346 with the application of 
a thorough cluster analysis. We first assess the topology of young stellar clustering in the 
region with the kernel density estimation technique, and the distribution function 
of stellar separations. We then decipher the clustering behavior of young stars in NGC\,346 
with the construction of their observed two-point correlation function, i.e., their autocorrelation 
function and its comparison to those from a series of simulated stellar distributions. 
We explore, thus, the limitations of this method and provide an accurate interpretation 
of the observed autocorrelation function. We, thus, carefully characterize 
the complex clustering behavior of young stars in NGC\,346.


The paper is organized as follows. 
In Section\,\ref{s:tpcf} we present the stellar surface density map
of NGC\,346, and the distribution function of stellar separations. In
Section\,\ref{s:acf} we present the autocorrelation function and discuss the first
observable evidence that NGC\,346 contains (at least) two components
with very distinct distributions. We compare the autocorrelation function 
of NGC\,346 with simulations of centrally condensed and self-similar stellar
distributions in Section\,\ref{s:acfint}. We show that indeed the
stellar distribution in the region is the result of two individual
stellar components, a central condensed and an extended fractal
distribution. In the same section we also constrain the basic parameters 
of these components using  dedicated simulations of mixed stellar 
distributions. Finally, in Section\,\ref{s:sum} we summarize our results and discuss
their implications to our understanding of clustered star formation.
Concluding remarks of our study are given in Section\,\ref{s:concl}.
Additionally, in Appendix\,\ref{s:appendix} we present our library of simulated
autocorrelation functions to be used for the interpretation of that observed
in NGC\,346, and in Appendix\,\ref{d3toeta} we provide an empirical calibration 
between three-dimensional and two-dimensional fractal dimensions over a 
wide range of  values.


\section{The Spatial Distribution of Young Stars}\label{s:tpcf}
\begin{figure}
\centerline{\includegraphics[angle=0,clip=true,width=\columnwidth]{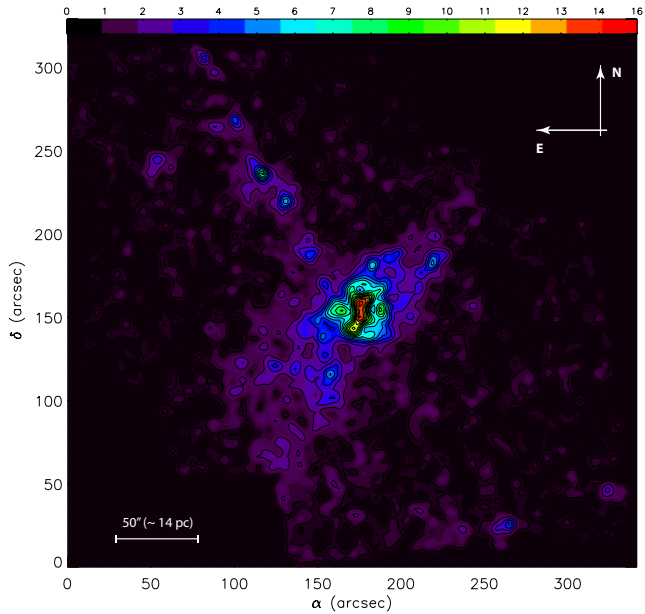}}
\caption{The surface stellar density (significance) map constructed from the sample of young stars in the region
of NGC\,346 with the use of the KDE technique and a 5\arcsec\ Gaussian kernel. Isopleths 
at different  density levels, drawn with different colors, signify the individual stellar clusterings 
identified in this region. The color-bar indicate different density levels in $\sigma$, the 
standard deviation of the measured stellar density in the whole region. North is up and East 
is to the left. Coordinates are given as in Figure\,\ref{f:map}.}
\label{f:kdemap}
\end{figure}

\subsection{The Kernel Density Estimation Map}\label{s:kde}

We construct the surface density map using the {\sl kernel density estimation} (KDE) 
technique \citep{Silverman92}. This technique smooths individual data point locations 
with Gaussian kernels to form a continuous spatial distribution. The fundamental
parameter is the full-width-at-half-maximum (FWHM) of the kernel.
For the purposes of our study the selection of this parameter was based on the desired 
`resolution', i.e.,  the smallest stellar structure to be identified. 

Based on the sizes of the detected stellar clusterings in Paper\,I, we
optimize the construction of the surface density map with a FWHM
$\sim$\,5\arcsec\ (equivalent to $\sim$\,1.5\,pc)\footnote{In 
Paper\,I we applied an unsupervised cluster detection,
based on the 20th nearest-neighbor density map of the region.}. The
produced KDE density map of NGC\,346 is shown in Figure\,\ref{f:kdemap}.
In this map small stellar clusterings, coinciding with the stellar
groupings previously detected using the nearest-neighbor method,
appear as over-densities at density levels of \gsim\,2$\sigma$. However, 
all these clusterings are part of a large stellar structure at the 1$\sigma$ 
density level.  At the lowest, $\sim1\sigma$ significance level, two large loose
stellar over-densities are apparent in the KDE map, which coincide
with features that can also be discerned in ionized gas and PAH
emission: 1) The central ``bar'' of the star-forming region, i.e., the
bright emission region, extending from southeast to northwest
\citep{rubio00}, and 2) the ``northern arc-like arm'' identified in
mid-infrared wavebands to extend from the center of the region to the
center-northeast part of the field \citep{gouliermis08}.

The KDE map of the region depicts, similarly to the nearest-neighbor map, that the region of NGC\,346 
includes several separate clusterings, defined by the 2$\sigma$ and 3$\sigma$ isopleths, 
characterized in previous studies as individual sub-clusters \citep[][Paper\,I]{sabbi07}. However, the KDE map illustrates in a far 
more meticulous manner another critical characteristic of the stellar clustering 
in NGC\,346. There are several compact stellar clumps, which do not appear to be independent, but they emerge as 
small over-densities within the NGC\,346 bar and northern arm. The bar itself is revealed as a large stellar structure (aggregate) 
in the KDE map at 2$\sigma$ significance.
Stellar distribution within this aggregate is organized in a segregated fashion with the small compact clusterings 
surrounding a central massive cluster, which appears circular at $\sim$\,4$\sigma$ significance  and higher (Figure\,\ref{f:kdemap}).  
Whether there is any hierarchy in the manner the aggregate is assembled, whether the 
central cluster is indeed centrally concentrated and how all the over-densities are connected to each 
other within the same large structure, are the questions we intend to explore with our analysis.

\begin{figure}
\centering
 \includegraphics[width=\columnwidth]{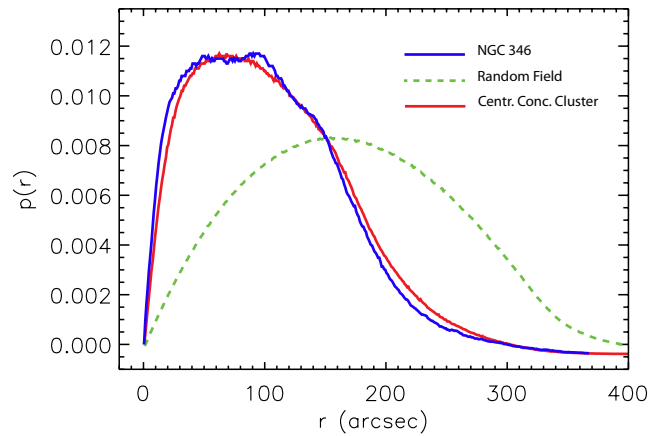} 
\caption{Distribution function for separations between the young stars  
in our sample for NGC\,346 (blue line). The separations distribution $p(r)$ 
for a simulated field of stars with uniform surface-density (i.e., random distribution; green-dashed line), 
and that for a centrally concentrated cluster with density gradient $\propto r^{-1.7}$ (red line) are 
plotted for comparison. The construction of both these artificial stellar distributions are discussed in detail in Appendix\,\ref{s:appendix}.
The separations distribution for NGC\,346 shows multiple maxima, and its shape is very close to that for the centrally concentrated 
cluster (see Section\,\ref{s:346ascluster}, for a discussion on this similarity).}
\label{f:pdf}
\end{figure}

\subsection{Stellar Separations Distribution Function}\label{s:pdf}

We derive the probability distribution function of stellar separations
for the stars in our sample, following the recipe of \cite{cw04}. 
The probability function, $p_i(r_j)$, for each star $i$ is calculated as
the number of pair separations $N_{ij}$ that fall in the separation bin 
centered on $r_j$ divided by the total number of separations:
\begin{eqnarray}\label{eq:pdf}
p_i(r_j) & = & \frac{2\,N_{ij}}{N\,(N-1)\,dr} \,,
\end{eqnarray}
where $N$ is the total number of stars. The 
distribution function $p(r)$ is calculated in every separation bin from 
the sum of $p_i(r_j)$: 
\begin{eqnarray}
p(r_j) & = & \sum_{i=1}^{N}p_i(r_j).
\end{eqnarray} 
The probability that the projected separation between two randomly 
chosen stars is in the interval $(r, r+dr)$ is given by $p(r)dr$.  

\begin{figure*}
\centering
\includegraphics[width=\textwidth]{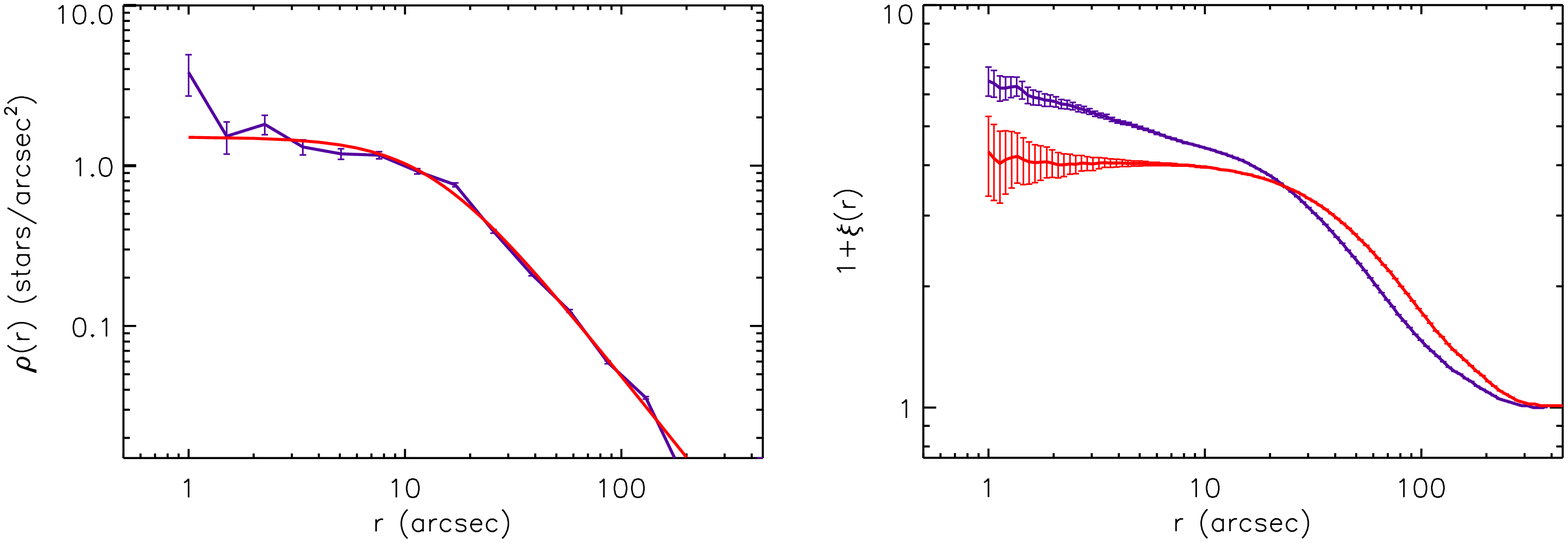} 
\caption{Left: Radial stellar surface density profile of the young stars in the observed field 
of NGC\,346 (in blue). This profile, constructed in concentric annuli, is being build for a first 
assessment of the structural parameters of the stellar cluster that would best 
represent the stellar distribution in NGC\,346. The best fitting profile model with 
core radius $r_c \simeq 15\arcsec$ and density profile slope $\gamma \simeq 1.7$ 
is plotted in red. Right:  The autocorrelation function of the young stars in NGC\,346 
(blue) and that of a simulated cluster (red) with the characteristics of 
the best-fitting model for the observed density profile. The differences between the 
two autocorrelation functions are striking, demonstrating that the clustering of stars in 
NGC\,346 cannot be explained as being produced by a single centrally concentrated 
cluster, in contrast to what the radial stellar density profile implies. The right panel is 
discussed thoroughly in Section\,\ref{s:onecluster}. It is shown here to demonstrate 
the disagreement in the results between the autocorrelation function and the density 
profile analyses.}
\label{f:densprof}
\end{figure*}

The separations distribution function for the young stellar population in 
NGC\,346  is presented in Figure\,\ref{f:pdf} (blue line). We also show the stellar 
separations distribution for a uniform (not clustered) stellar
field (green-dashed line), and that for a simulated centrally concentrated cluster
(red line). Details on the simulations are discussed in Appendix\,\ref{s:appendix}.
Perusal of Figure\,\ref{f:pdf} shows that overall the separation
distribution compares quite well with that of the centrally concentrated
cluster with surface density profile $n \propto r^{-1.7}$. However, 
this agreement is surprising, since evidently both the 
observed stellar chart (Figure\,\ref{f:map}) and surface density map 
(Figure\,\ref{f:kdemap}) exhibit more structures than can be explained by a
simple spherical cluster. This similarity is further discussed in terms of the radial 
stellar density profile in Section\,\ref{s:346ascluster}.


The observed probability distribution also displays a number of local
maxima, which indicates the existence of some preferred length-scales 
in NGC\,346, perhaps resulting from multiple clustering in the region. The
clearest local maxima occur at separations of about 50\arcsec and
100\arcsec ($\sim$\,14\,pc and 28\,pc respectively). The smaller of
these scales coincides with the size of a central stellar
concentration, as revealed in the KDE map of Figure\,\ref{f:map} at
3$\sigma$ significance. The maximum at the larger length-scale concurs
with the average size of the stellar aggregate in the NGC\,346 bar (at
2$\sigma$ in the KDE map).

\subsection{Describing NGC\,346 as a single condensed cluster}\label{s:346ascluster}

Based on the stellar separation distribution function, presented in the previous
section, one might be tempted to interpret the entire stellar
distribution of NGC\,346 as originating from a single, centrally concentrated
cluster, in spite of the obvious asymmetries in the stellar map. We
performed a typical analysis by constructing the radial
surface density profile of young stars in the entire observed area, 
centered on the peak surface density. This profile is shown in blue 
in Figure\,\ref{f:densprof} (left panel). We then fitted to the observed profile 
a model cluster density profile with the functional form 
prescribed by \citet[][see Appendix\,\ref{s:effclusters}]{eff87}. 
The best-fitting model is also shown in Figure\,\ref{f:densprof} 
(left) in red. 

Considering that NGC\,346 is certainly not a single condensed cluster,
its radial profile was surprisingly well reproduced by 
models for clusters with core radii of $r_c \simeq 14.9\arcsec \pm 0.7\arcsec$, 
and density profile slopes of $\gamma \simeq 1.69 \pm 0.06$. Such a model
cluster is used for comparing the stellar separations distributions shown in 
Figure\,\ref{f:pdf}. This agreement shows that caution is warranted 
when interpreting radial density profiles, as well as stellar separations distribution 
functions, in particular in cases where the observations do not allow to recognize 
prominent asymmetries. It demonstrates the need for a diagnostic tool 
for stellar distributions, which is equally sensitive to structure afar from the 
regions with the highest stellar density. To this end we use the two
point correlation (autocorrelation) function. The diagnostic power of this function is
demonstrated in Figure\,\ref{f:densprof}, right panel, which shows that
the best-fit cluster distribution fails completely to reproduce the observed
autocorrelation function (see Section\,\ref{s:onecluster}), in spite of the 
excellent fit to the radial profile (Figure\,\ref{f:densprof} left) and stellar separations 
distribution function (Figure\,\ref{f:pdf}).

\section{The Autocorrelation Function of Young Stars in NGC\,346}\label{s:acf}


The degree of clustering of stars can be quantified by using the two-point correlation function 
\citep[][Section\,45]{peebles80}. Applied to stars in the same sample, this function becomes an {\sl autocorrelation
function} (ACF). In the following, we broadly follow the method introduced by \cite{peebles80} for cosmological 
applications and modified by \cite{gomez93} for characterizing the clustering behavior 
of T\,Tauri stars in Galactic star-forming regions. Other recent investigations applied this method to 
observed samples of star clusters in remote galaxies \citep[e.g.,][in the Antennae and M\,51 galaxies 
respectively]{zhang01, scheepmaker09}. 


The ACF is defined as:
\begin{eqnarray} \label{eq:autocorrelation}
1+\xi (r) & =  & \frac{1}{\bar{n}N}\sum_{i=1}^{N} n_{i}(r),
\end{eqnarray} 
where $n_{i} (r)$ is the number density of stars found in an
aperture of radius $r$ centered on, but excluding star $i$. $N$ is 
the total number of stars and $\bar{n}$ is the
average stellar number density. The corresponding uncertainties 
based on \cite{peebles80} are given by:
\begin{eqnarray}\label{eq:acferr}
\delta(r) & = & \sqrt{N} \cdot\left(\frac{1}{2}\sum_{i=1}^{N} n_{\mathrm{p}}(r)\right)^{-1/2}, 
\end{eqnarray}
where $n_{\mathrm{p}}(r)$ is the number of pairs formed with the
central star $i$ of the current aperture, and the factor $1/2$ accounts for not counting every pair twice.

In general, $\xi (r)$ is
defined such that $\bar{n}[1+\xi(r)] d^2r$ is the probability of
finding a neighboring star in a area of radius $r$ from a random 
star in the sample. This means in effect that 
$1+\xi (r)$ is a measure for the mean surface density within radius 
$r$ from a star, divided by the mean surface density of the total sample 
(i.e. the surface density enhancement within radius $r$ with respect to 
the global average). Therefore, for a random stellar distribution $1+\xi(r) = 1$, 
while for a clustered distribution  $1+\xi(r) > 1$. 

 For a hierarchical, or fractal, distribution of stars the ACF yields a 
 power-law dependency with radius of the form 
$1+\xi (r) \propto r^{\eta}$ \citep{gomez93}. For such a distribution, the total number of 
stars $N$  within an aperture of radius $r$ increases as $N\propto r^{\eta}\cdot r^{2}  = 
r^{\eta+2}$. The power-law index $\eta$ is related  to the two-dimensional 
fractal dimension $D_{2}$ as $D_{2} = \eta +2$ \citep{mandelbrot83}. Power-law ACFs 
have been observed for interstellar gas over a large range of 
environments, and have been interpreted as indications of hierarchical structuring of the
gas. The derived typical (three-dimensional) fractal dimension of $D_{3} \sim2.3$ 
\citep[e.g.][]{elmegreen96, elmegreen01b} apparently comes from the same underlying
distribution as found for extragalactic star-forming regions in NGC\,628 with a two-dimensional 
fractal dimension of $D_{2}\simeq 1.5$ \citep[][]{elmegreen06}. However, the conversion 
$D_{2} = D_{3}-1$ is not generally valid is discussed in \cite{elmegreen96} and \citep[][]{elmegreenscalo04}. 
In Appendix\,\ref{d3toeta} we calibrate this relation over a large range of fractal dimensions. 


\subsection{The Effect of Limited Observed Field-of-View}\label{s:edgeeffect}

\begin{figure}
\centering
 \includegraphics[width=\columnwidth]{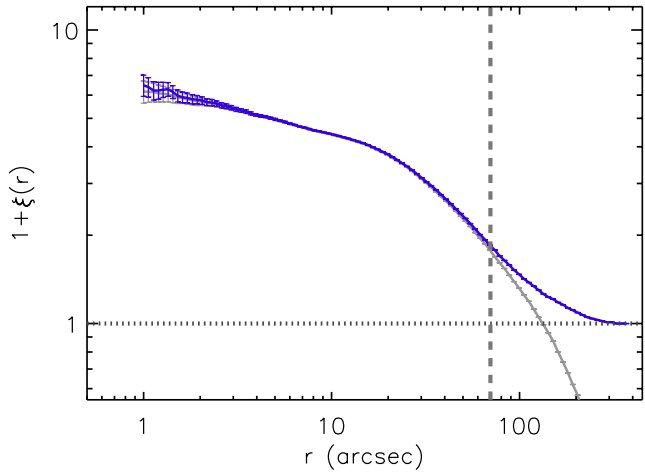} 
\caption{The ACF of the young stellar population (PMS and UMS stars) 
in NGC\,346. The horizontal dashed line corresponds to the ACF of a random
field and equals unity. The grey line corresponds to the ACF uncorrected for
edge effects, while the blue line depicts the corrected ACF (see discussion 
in Section\,\ref{s:edgeeffect}.}
\label{fig:autocorrelation}
\end{figure}

The ACF, calculated according to Eq.\,(\ref{eq:autocorrelation}), is
sensitive to the size of the surveyed area, because apertures around stars close to the edge 
fall partly outside the survey area, measuring a too low average stellar surface density. 
This introduces a very steep decrease, dropping well below unity for larger
separations due to missing stars outside of the observed field-of-view.  
This behavior is demonstrated in the ACF of NGC\,346 shown with a grey line in Figure\,\ref{fig:autocorrelation} 
(described in Section\,\ref{s:acf346}). The correct way of dealing with this issue is by ``masking'' the apertures,
i.e., by dividing only the part of each aperture that overlaps with the observed field-of-view when calculating 
average surface densities ($n_{i}$ in Eq.\,\ref{eq:autocorrelation}). With this correction the ACF drops smoothly to the level of the 
random field for larger separations (blue line in Figure\,\ref{fig:autocorrelation}). From Figure\,\ref{fig:autocorrelation} we 
assess that the correction for the ACF of NGC\,346 becomes dominant at separations larger than $\sim$\,70\arcsec.
We consider this ACF, as well as all ACFs simulated in our analysis, to be reliable for separations up to this limit.


\subsection{The Observed ACF of NGC\,346}\label{s:acf346}

The ACF for the young stellar population of NGC\,346, constructed 
according to Eq.\,(\ref{eq:autocorrelation}), is plotted with respect to projected stellar separations 
in Figure\,\ref{fig:autocorrelation}. Error bars are determined according to  
Eq.\,(\ref{eq:acferr}). The clustering of young stars in 
NGC\,346 becomes stronger, i.e., larger values of $1+\xi(r)$, at smaller stellar separations. 
This figure also shows that the stellar clustering in NGC\,346 changes behavior at different 
scales. Two distinct parts in the ACF plot can be distinguished at the separation of about 20\arcsec\  
($\sim$\,5\,pc). Both parts show an almost linear decrease of the ACF with radial distance, but with
significantly different slopes. Each of these power-law dependencies is similar to that expected for 
a fractal stellar distribution. However, a purely hierarchical stellar distribution exhibits a single-slope 
increase of $1+\xi$ with smaller separations over all scales unlike the ACF of Figure\,\ref{fig:autocorrelation}
that shows a clear break.  

We verify that the ACF of young stars in NGC\,346 is well-described by a broken power-law, and we determine 
its slopes, $\eta$, by fitting such a power-law function. We establish the indexes of the two 
corresponding linear parts by applying a  Levenberg--Marquard nonlinear least square minimization technique \citep{levenberg44, 
marquardt63}, as implemented in IDL by \cite{markwardt09}. The two power-law slopes in the ACF, as well as the position of the 
break point along the abscissa are the free parameters in our fit. The break in the slope occurs at separations of 
20.85\arcsec\ ($\sim$\,5.8\,pc). The 
inner part (r \lapprox\,21\arcsec) has a power law index $\eta_{\rm in} \simeq -0.18$, which corresponds 
to a 2D fractal dimension of $D_{2} \simeq 1.8$. For separations 21\arcsec \lapprox\ r \lapprox\,70\arcsec\ (6\,pc 
\lapprox\ r \lapprox\,20\,pc) the ACF has a power-law index of $\eta_{\rm out} \simeq -0.58$ ($D_{2} \simeq 1.4$). As we discussed earlier, 
separations beyond the limit of $\sim$\,70\arcsec\ cannot be considered in our analysis, since at these separations 
the ACF correction for the finite observed field is dominant. 

The fractal dimension found for the small stellar separations in NGC\,346 is quite close to the geometrical 2D dimension, 
suggesting a smooth stellar distribution at these scales. It is interesting to note that at larger scales the derived 
smaller fractal dimension suggests a more clumpy distribution of stars. This fractal dimension of $D_{2} \simeq 1.4$
agrees very well with that found by several authors for the interstellar gas in the Milky Way and other galaxies 
\citep[e.g.,][]{falgarone91, westpfahl99, kim03}, as well as with that found for stellar clusters in external galaxies \citep[e.g.,][]{zhang01, 
elmegreen06, scheepmaker09}, and for young stars in several dwarf galaxies \citep{odekon06}. These investigations argue 
that the derived fractal dimensions originate from the turbulent motions in the ISM. 

Projected fractal dimensions with values between $D_{2} = 1.3$ and 1.5 are considered to be consistent with analogous 
measures of the 3D fractal dimension of $D_{3} \sim 2.4$ derived by both fractional Brownian motion simulations 
\citep[see, e.g.,][]{stutzki98} and from laboratory or numerical turbulent flows \citep{mandelbrot83, sreenivasan91, federrath09}.  
It should be noted that this argument is valid only when the simple conversion $D_3=D_2+1$ holds. Based on our simulations of 
self-similar stellar distributions (see Appendix\,\ref{s:fractaldistr}), we show that
{\em the conversion $D_3=D_2+1$ is not generally valid} (see Appendix\,\ref{d3toeta}). We also establish an empirical 
relation between the original $D_3$ used for constructing the distributions and the $D_2$ values derived with the use of 
the ACF. 

Our findings on the ACF of NGC\,346 clearly suggest that stellar clustering at scales larger than $\sim$\,21\arcsec 
is hierarchically driven by turbulence and thus related to the structure of the ISM. Our simulations of 
fractal distributions, discussed in Appendix\,\ref{s:fractaldistr}, show that indeed hierarchical stellar 
clustering produces a linear dependency of ACF in log-log, but this dependency extends with the same index $\eta$ 
across the whole considered length-scales range. Cases where 
this has been observed are reported by \cite{gomez93}, \cite{larson95}, and \cite{simon97} in star-forming 
regions of the Milky Way. The fact that the ACF of Figure\,\ref{fig:autocorrelation} behaves like a broken power-law 
implies that the stellar distribution in NGC\,346 is neither a purely hierarchical, nor a pure centrally condensed distribution.
In fact, the clear change of slope in the observed ACF provides a strong indication that 
there are multiple components present, whose distribution is quantitatively and significantly different. 
The first, high density, component -- e.g., a centrally concentrated cluster -- affects preferentially the most 
populated regions of the observed field, while the second, more extended component has significant members 
throughout the region. This is consistent with the observation that the bar of NGC\,346 is a loose stellar congregate, 
with a definitive compact stellar concentration at its center (Figure\,\ref{f:kdemap}).  





\begin{figure}
\centering
 \includegraphics[width=\columnwidth]{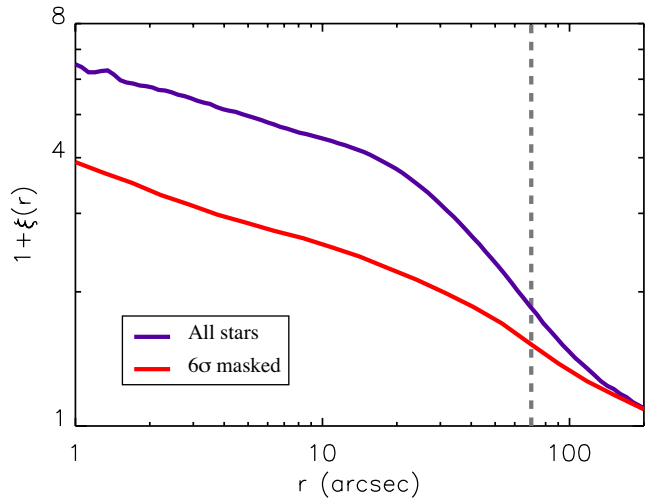} 
\caption{The ACF of the young stellar population in NGC\,346 through its complete extend (in blue) 
compared to the ACF of the same population with the central compact region masked. The ACF derived 
after masking the stellar sample at 6$\sigma$ KDE-significance is plotted in red. This ACF 
exhibits  a single slope in contrast to the broken power-law observed for the whole 
sample. This supports the interpretation that the stellar distribution in NGC\,346 is governed by two
components, a compact and an extended one. The  vertical dashed line indicates the 
separation of 70\arcsec\ beyond which the effect of finite field-of-view becomes dominant.}
\label{f:cclusmask}
\end{figure}

We test the two-components hypothesis by a simple experiment. We repeat our calculation of the ACF while 
excluding the central compact part of the observed field. Specifically, we mask the stellar sample at significance 
levels varying between 3 and 8$\sigma$ from the KDE map of NGC\,346 (Figure\,\ref{f:kdemap}). We find that for values 
between 5 and 7$\sigma$ the resulting ACF is very close to a single power law, with an exponent of $\sim -0.2$.
The derived ACF for masking value of 6$\sigma$ is plotted in red in Figure\,\ref{f:cclusmask}.  For higher values of $\sigma$, 
too many stars in the central compact part are retained and the corresponding ACF still exhibits the characteristic break.
This experiment provides more confidence that indeed NGC\,346 is the conjugated result of two different stellar distributions. 
The following section is dedicated to obtaining quantitative constraints on the nature of these separate components through 
numerical simulations.



\section{Numerical Simulations}\label{s:acfint}

We derive the nature of the components that dominate the stellar clustering by performing a 
thorough set of numerical simulations. These simulations are essential for the correct interpretation 
of observed ACFs, and quantify the uncertainties of the derived clustering parameters. In Appendix\,\ref{s:appendix} 
we present a library of ACFs on the basis of simulations of centrally condensed and self-similar stellar distributions.
These simulations aid the interpretation of the ACF assuming a {\em single} type of clustering (compact or fractal). 
Here we discuss simulations directly applicable to NGC\,346, which shows a  more complicated clustering behavior
with the co-existence of {\em multiple} stellar components. All considered simulations contain the same 
number of stars as observed in NGC\,346 in comparable field-of-view. This implies that an applicable 
simulation should reproduce not only the shape of the observed ACF but also the absolute values. 

\subsection{Simulations of a Single Centrally Condensed Cluster}\label{s:onecluster}

In Section\,\ref{s:tpcf} we found that both the stellar separations distribution function (Figure\,\ref{f:pdf})
and the stellar surface density profile of young stars in NGC\,346 (Figure\,\ref{f:densprof}) imply that
the young stars are mostly distributed in a centrally condensed fashion. Therefore, the simplest possible 
model to describe the ACF of NGC\,346 would consist of a single, centrally concentrated cluster.
In Appendix\,\ref{s:ccclusters}, among the considered types of centrally condensed cluster models, 
that proposed by \citet[from hereon the EFF model]{eff87} is the most appropriate, since it refers to non-tidally 
truncated clusters like those found embedded in star-forming regions. 

We simulated an EFF cluster whose density profile fits best that of NGC\,346 (shown in Section\,\ref{s:346ascluster}) 
and constructed its ACF. We compare the ACF of NGC\,346 to that of this cluster in Figure\,\ref{f:densprof} (right panel). 
In this figure, it can be seen that while the density profiles fit very well, the corresponding ACFs are strikingly 
different from each other. At smaller separations the trend of 
$1+\xi(r)$ cannot be reproduced by the ACF of the EFF cluster, because it is flat. The decrease of $1+\xi$ with separation for larger scales with index $-0.4$ 
does not reproduce that found for NGC\,346 ($-0.6$; Section\,\ref{s:acf346}) either. This result highlights the fact
that while the surface density profile of a stellar concentration may give the impression of a centrally concentrated 
cluster, its ACF may clearly indicate that this is not the case. As a consequence the use of surface density 
profiles for characterizing uncertain stellar distributions (e.g., open clusters and loose stellar groups) should be 
made with caution, as it may not represent reality.



\begin{figure}
\centering
\includegraphics[width=\columnwidth]{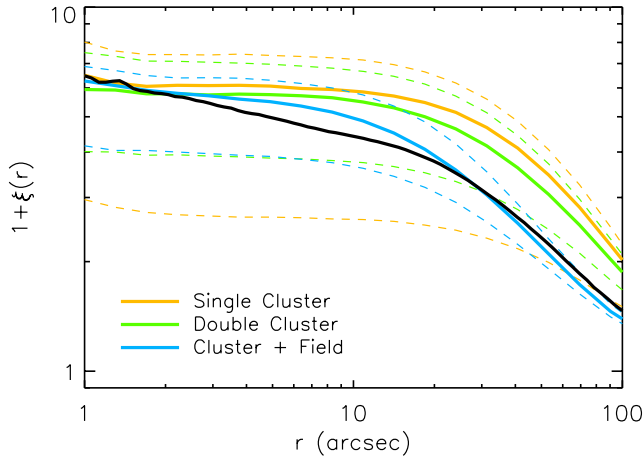} 
\caption{The `most successful' ACFs of stellar distributions, constructed according 
to the scenarios discussed in Sections\,\ref{s:onecluster}\,-\,\ref{s:oneclusterandfield}. 
We show the results for a single cluster (orange line), a double cluster (green line) 
and a cluster in a homogeneous field (blue line). They all fail to explain the observed 
ACF (shown with the black line). We demonstrate the spread in these simulations by
also showing, in dashed lines, the results after changing the density profile index of the 
considered clusters by $\gamma = \pm 0.2$. Even the `best' simulations of those that do 
not include a self-similar stellar component fail to reproduce both the shape and absolute 
values of the ACF of NGC\,346.}
\label{f:wfitacf}
\end{figure}

We explored the possibility that other EFF clusters may reproduce the observed ACF, but  
we found that {\sl all} simulated EFF profiles fail to reproduce the main characteristics of the observed 
ACF. They all produce constant $(1+\xi)$ values at scales smaller than their core-radii, not 
agreeing with the observed ACF of NGC\,346 at these scales. 
In particular, those clusters that had stellar densities at shorter separations compatible to that of 
NGC\,346 contained unrealistically large numbers of stars within their core radii. 
 The ACF of the ``most successful'' set of our simulations of 
a single centrally condensed cluster is plotted in Figure\,\ref{f:wfitacf} (orange line), where it is 
shown that the single-cluster scenario cannot reproduce successfully the observed ACF. From this analysis  
we conclude that while the stellar density map of NGC\,346 indicates the existence 
of a condensed stellar cluster, the ACF indicates that not all young stars in NGC\,346 can 
belong to the central cluster and thus another, more extended, component should be also considered. 
This result is additionally supported by the decomposed ACF of NGC\,346 (Figure\,\ref{f:cclusmask}), 
constructed by masking the central condensed cluster

In the following sections we explore the constraints we can obtain on the nature of this extended 
component. We identify the type of this component by applying simulations of the combination of 
two individual stellar 
distributions. The total number of stars and the field-of-view in these simulations are again identical
to the observed ones in NGC\,346. The basic input parameters in our experiments are (1) the fraction
of stars with respect to the total considered number that belong to each component, (2) the core radii and 
$\gamma$ indexes of each considered condensed cluster, and (3) the fractal dimensions 
of the simulated self-similar distributions.  

\subsection{Simulations of two Centrally Condensed Clusters}\label{s:twocluster}

We study the possibility of reproducing the main characteristics
of the observed ACF, assuming two centrally condensed stellar components. This is motivated
by the fact that in the stellar surface density (KDE) map of Figure\,\ref{f:kdemap} a small number ($<$\,4) of
isolated density peaks can be seen, depending on the chosen significance level. We produced mock 
stellar catalogs with the addition of two EFF clusters, with a total number of observed stars equal to 
that of NGC\,346 (5,150 stars). We produced a grid of combinations in the stellar distribution by varying 
the following parameters: 1) Projected separation between the clusters, which varied from 10\arcsec\,to 
25\arcsec, 2) the fraction of stars belonging to the first cluster (0.5 to 0.7), 3) the EFF profile indexes 
$\gamma$ (between 1.6 and 2.2), and 4) the core radii, $r_c$, which was set between 6\arcsec\,and 12\arcsec. 

The best correspondences between models and observations were found for low values of $\gamma$ ($\sim$1.8) and core radii 
($\sim$\,8\,-\,10\arcsec). Distributing the stars in roughly equal amounts over two clusters allows 
for somewhat smaller individual clusters, which alleviates to some extent the problem of having too
many stars within the core radius -- and thus absolute ACF values higher than the observed. 
 This is demonstrated in Figure\,\ref{f:wfitacf}, where the most well representative ACF from 
our simulations of two centrally condensed clusters is shown (in green). If the 
separation between the clusters is tuned to be roughly consistent with the observed break of 
about 21\arcsec, the smaller clusters combined can still yield a weak break at this separation in the ACF,
but cannot reproduce the power-law trend of the observed ACF at scales below 15\arcsec. 
In general, all of the considered combinations of two EFF clusters completely failed to reproduce 
the behavior of the ACF of NGC\,346 in the complete range of separations. 

\subsection{Simulations of a Single Cluster plus a Homogeneous Field}\label{s:oneclusterandfield}

One of the striking features of the distribution of the young population in NGC\,346 is that there are 
stars found distributed throughout the observed region (see also Paper\,I), which may be an indication 
of a truly dispersed population, i.e., a (young) background field population. We simulate thus clusters that are 
located within a homogeneous field. In these simulations we vary the following parameters: 1) The fraction of stars
that belong to the EFF cluster ($f_{\rm cl}$: 0.4 to 0.7), 2) the core radius (3\arcsec\,-\,15\arcsec) and 3) the 
slope $\gamma$ (1.5\,-\,3.0) of the cluster profile. 

The best agreement with the observed ACF is found for
$f_{\rm cl} \simeq 0.5$, core radii between 8\arcsec\,and 10\arcsec and quite high $\gamma$ 
values from 2.2 to 2.5. For these parameters the peak stellar surface density (i.e., the ACF values 
at the smallest separation bin) and the ACF at separations larger than 21\arcsec\, are well reproduced. 
However, all simulated distributions fail to reproduce the ACF behavior at separations smaller
than 15\arcsec, in identical manner as the two previous scenarios (see Figure\,\ref{f:wfitacf}). 
The origin of this general mismatch with the observations is the fact that the stellar surface density of both 
the underlying EFF distributions and the random field does not increase at continuously smaller 
spatial scales. 

\begin{figure}
\centering
\includegraphics[width=\columnwidth]{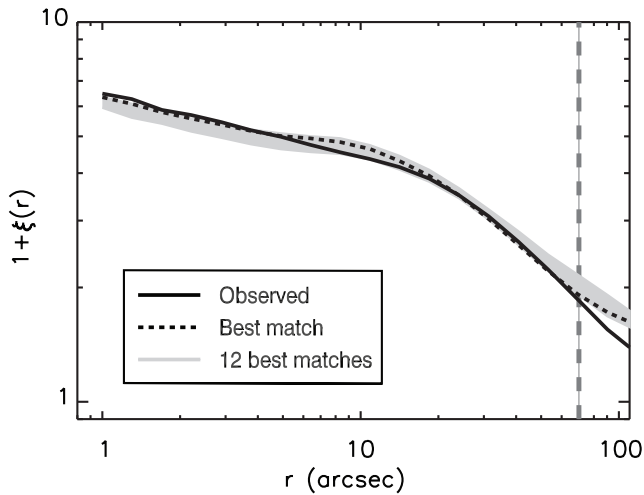} 
\caption{The observed ACF of NGC\,346 (solid black line) with the area filled by the ACFs of the best-representative 
simulated mixed distributions that assume a central compact stellar component and an extended fractal 
one (light grey). The useful separations range of \lsim\,70\arcsec, where the ACF indexes 
were determined, is indicated by the vertical dashed line. These models succeed in reproducing the 
observed ACF at the shortest and longest stellar separations. They have, though, the tendency to
over-predict the ACF values at separation of $\sim$\,10\arcsec.
The ACF of the best-matching model is plotted with a short dashed line. 
These simulations provide constraints for the basic parameters of the contributing stellar distributions,
with  the central cluster having a core radius of $\sim$\,9\arcsec ($\sim$\,2.5\,pc) and extending with a slope 
$\gamma \sim$\,2.27, and the self-similar distribution having a fractal dimension of $D_{3} \sim$\,2.3.}
\label{f:bfitacf}
\end{figure}

\subsection{Simulations of a purely Self-Similar Distribution}\label{s:purefractal}
 
In Appendix\,\ref{s:fractaldistr} we present a detailed account of the behavior of the 
ACF for self-similar, i.e. fractal, stellar clusterings, based on artificial stellar distributions.
It is shown that hierarchical stellar distributions are characterized by a monotonous (i.e.  
single-slope) linear dependency of the ACF on stellar separations in log-log, which extends
across the whole considered length-scales range. The shape of the observed 
ACF for NGC\,346 (Figures\,\ref{fig:autocorrelation} and \ref{f:cclusmask}) is proven to behave 
like a broken power-law with different slopes for different separations ranges (Section\,\ref{s:acf346}). 
As a consequence, with our simulations of pure fractal stellar distributions we are able to reproduce 
each one of the two linear parts of the observed ACF, but not its complete shape across the whole separations 
range. As we discuss in Section\,\ref{s:acf346}, and show in the following section, a second stellar 
component must be considered in order to successfully reproduce the broken power-law behavior of this ACF.

\subsection{Simulations of a Single Cluster plus a Self-Similar Distribution}\label{s:oneclusterandfractal}

The decomposed ACF of Figure\,\ref{f:cclusmask}, constructed by masking the central condensed 
stellar concentration of the region, strongly suggests the existence of an underlie self-similar distribution 
of young stars in NGC\,346, and it provided the first evidence that the system may be the combined result of 
two such diverse stellar distributions. Indeed, we find that the most successful simulations in replicating the 
observed behavior of ACF for NGC\,346 are those which contain both a centrally concentrated dense cluster 
and a large-scale hierarchical (fractal) distribution. 


We constructed a sample of distributions by varying the following
parameters. (1) The fraction of stars that belong to the cluster
$f_{\rm cl}$ (between 0.3 and 0.7), the core radius of the EFF cluster
(7\arcsec\,-\,12\arcsec), the $\gamma$ index of its profile (1.8\,-\,3.0) and the
three-dimensional fractal dimension $D_3$ of the fractal distribution 
(from 2 to 2.6). There is only a small subset of these models that is able to 
reproduce satisfactorily that observed ACF, i.e. the peak value of $1+\xi$, the 
power-law slope for separations between 1\arcsec\ and 21\arcsec, and the 
steeper slope for larger separations (21\arcsec\,to 70\arcsec). The simulations 
that represent best the observed ACF have 2,300 to 2,500 stars in the cluster, 
which has a core radius of 4\arcsec\ to 9\arcsec\, 
and $\gamma$ between 2.25 and 2.35. The best-matching extended
component has a high fractal dimension (2.2 $<$ $D_3$ $<$ 2.4).
Smaller clusters systematically over-predict the ACF peak value (at the shortest-scale 
bin). Clusters with core radii \gsim\,10\arcsec fail to simultaneously reproduce the 
AVF behavior at both small and large separations. They produce too much 
correlation for large separations, which can be remedied by increasing the
fraction of stars in the cluster but then the power-law behavior at shorter spacings 
is lost.

With these dedicated simulations we succeeded in reproducing both the shallow linear 
behavior of NGC\,346 ACF at short separations, as well as its steeper drop 
at large separations. We reproduced the ACF index for every part of its broken power-law, as well
as the separation limit, where the break occurs. Twelve of our combined simulations reproduce  
the ACF of NGC\,346, providing constraints to the basic parameters of the assumed stellar distributions.
In all these simulations the core radius of the central cluster is found to be practically unchanged and equal to
$\sim$\,8\arcsec\,-\,9\arcsec. The fraction of stars in the EFF cluster varies around 0.4, and the slope $\gamma$ 
of the assumed cluster profile has values $\gamma\simeq$\,2.20 to 2.35. Finally, the input three-dimensional 
fractal dimension of the best-fitting simulations varies at values between $\sim$\,2.20 and 2.36. 

\begin{figure}
\centering
\includegraphics[width=\columnwidth]{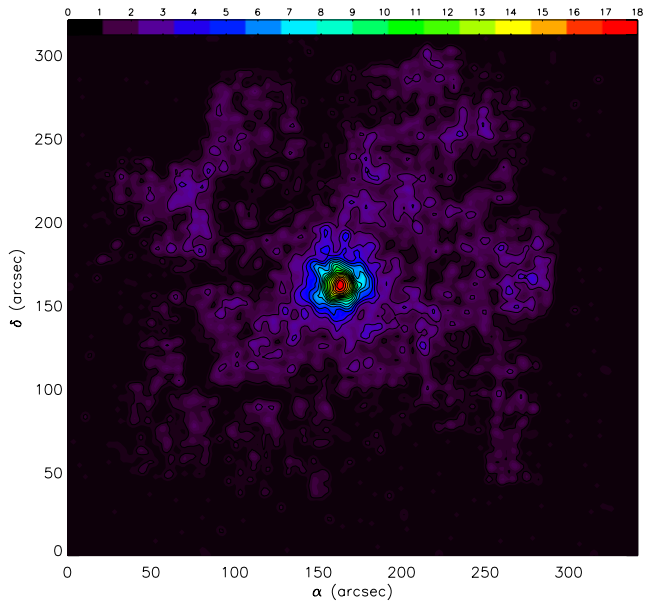} 
\caption{Surface stellar density (significance) map constructed from the synthetic stellar distribution 
that represents best the stellar clustering in NGC\,346.
The map is constructed with the use of the KDE technique and a 5\arcsec\ Gaussian kernel,
in comparison to the NGC\,346 map of Figure\,\ref{f:kdemap}. The color-bar indicates different density levels in $\sigma$, the 
standard deviation of the measured stellar density in the regions. The artificial field-of-view is  
identical to the observed field-of-view. Coordinates are given as in Figure\,\ref{f:kdemap} for NGC\,346.}
\label{f:kdemodel}
\end{figure}

The observed ACF of NGC\,346 and those of the twelve most successful simulations are shown in Figure\,\ref{f:bfitacf}.
It should be noted that the modeled ACFs are very sensitive to the chosen parameters. We established the best-matching 
models by iteratively refining the input parameters. In Figure\,\ref{f:kdemodel} the KDE surface density map of the most successful 
distribution (short dashed line in Figure\,\ref{f:bfitacf}) with $f_{\rm cl} =$\,0.4, $r_{\rm c} =$\,9\arcsec, $\gamma = 2.3$, and $D_{3}=$\,2.32, 
is shown to be compared to the KDE map of NGC\,346. This density map shows secondary peaks, resembling concentrations 
seen in NGC\,346. Nevertheless, we could not reproduce the amplitude of the second most 
important stellar clump of NGC\,346, which may mean that there is a second (much smaller) compact cluster
in the region. 

 
 

\section{Summary and Discussion}\label{s:sum}

In this study we present a thorough analysis of the clustering behavior of young stars in the star-forming region NGC\,346.
In particular, we construct the KDE density map and the separations distribution function, and we determine the ACF of the 
rich sample of low-mass PMS stars  (supplemented by the young massive stars) in the region. We show that the ACF is a 
robust method to characterize the clustering of stars, providing that its behavior is accurately understood through dedicated 
simulations. We characterize the stellar clustering in NGC\,346, and interpret its ACF, through the construction and study of 
the ACFs of simulated centrally condensed and fractal stellar distributions. 

We previously found indications that NGC\,346 includes multiple compact stellar over-densities, and that the distribution of 
PMS stars in the region is hierarchical (Paper\,I). With our analysis here we demonstrate that the young stellar design of NGC\,346 is far more 
complicated. The comparison of the observed ACF with those of simulated stellar distributions shows clear evidence 
that NGC\,346 includes a hierarchical stellar component, extended across the whole field-of-view, and a compact stellar cluster, 
which appears more dominant in the central part of the observed region. 


The observed ACF of young stars in NGC\,346 is successfully reproduced if we assume that $\sim$\,40\% of the stellar population 
belongs to a condensed cluster with core radius of 
$r_{\rm c} \sim$\,2.5\,pc and a density profile index $\gamma \sim$\,2.3, ``embedded'' in a self-similar stellar distribution 
with a fractal dimension of $D_{3}\sim$\,2.3. NGC\,346 is, thus, a very interesting example of composed stellar distributions,
on length-scales typical for molecular clouds, providing an unprecedented insight of the topology of star formation. The fact 
that we still observe self-similar stellar clustering even after $\sim$\,3\,Myr is also interesting, because this time-scale provides a 
constraint on how long it takes for the stars to lose their primordial spatial distribution. 

The derived three-dimensional fractal dimension of $D_3 \simeq 2.3$, for the self-similar stellar component, fits very well to  
the value derived from numerical experiments of supersonic isothermal turbulence \citep{federrath09}, and 
with measurements inferred from observations of the interstellar gas \citep{elmegreen01b}. This agreement implies that the 
self-similar stellar distribution in NGC\,346 is  possibly inherited by the turbulent interstellar gas of the natal molecular cloud. 
 In our study for the determination of $D_3$ we use the ACF, \cite{federrath09} argue that the $\Delta$-variance \citep{stutzki98, 
ossenkopf08} is the most reliable method, and \cite{elmegreen01b} use the size distribution function of stellar aggregates.
Although different, all three studies agree in the derived value of $D_3 \sim$\,2.3.

 On the other hand, recent numerical simulations by \cite{girichidis12} and \cite{dale13} find a higher degree of hierarchy for 
stars formed from turbulent molecular clouds with smaller values of $D_3 \sim$\,1.6. These studies, which 
apply the $\cal{Q}$-parameter \citep[][]{cw04}, find similar fractal dimensions with that inferred from the same method for the Taurus 
molecular complex ($D_3 \sim$\,1.5), in line with the fractal dimensions of the ISM in this region \citep[see, e.g.,][]{alfaro11}. 

While the differences in the derived fractal dimensions may reflect discrepancies in the methods used, the may as 
well demonstrate the fact that stars  have a different spatial distribution to the gas from which they formed, or that the gas is not
similarly structured everywhere. Observations of nearby Galactic star-forming regions show that while the gas filaments have a 
smooth, radially decreasing density profile \citep[e.g.,][]{arzoumanian11}, the distribution of recently-formed stars 
($\tau \sim$\,1\,Myr) has a range of different morphologies, some being fractal and others centrally concentrated \citep{cw04}.

Whether or not every stellar distribution forms with substructure has yet to be determined. 
N-body simulations  suggest that even a moderate amount of dynamical interactions will partly erase substructure, 
forbidding a star-forming region from retaining a strong signature of the primordial ISM distribution \citep[e.g.,][]{scally02, gw04}. 
Considering that subsequent substructure is difficult to be formed, and dynamics may have aided the formation of a condensed 
cluster in NGC\,346, the observed substructure may be an upper limit of the primordial value.





The existence of the central compact cluster naturally complicates somewhat the star formation picture in the region.
Assuming that the clustered stellar population is coeval with the extended hierarchical stellar component, one has to
consider two scenarios for its formation: Either (1) this cluster was formed as a distinct prominent compact concentration 
amid a distributed stellar population, or (2) it was originally formed as part of the extended stellar component, and soon 
became centrally condensed due to rapid dynamical evolution. 

The first scenario implies the co-existence 
of two different modes of star formation; one that produces a centrally concentrated cluster plus another that 
is responsible for the distributed PMS stars. The formation of both clustered and distributed 
populations of young stars in a single molecular cloud is numerically described by \cite{bonnell11}, where 
both modes are determined by the local gravitational binding of the cloud. The bimodal star formation scenario 
is further supported by the large number of `unclustered' PMS stars (see also Paper\,I).

The second scenario assumes that the centrally condensed cluster in NGC\,346 is the merging product 
of distinct compact newly-born sub-clusters within the natal cloud. This scenario is supported by hierarchical 
fragmentation of the turbulent molecular cloud, which forms stars in many small sub-clusters \citep{klessenburkert}. These 
sub-clusters will interact and merge to form a dominant stellar cluster through closer and more frequent dynamical interactions 
\citep[][]{bonnell03}. The existence of compact PMS sub-clusters along the whole extend of NGC\,346 makes this scenario 
favorable. Recent N-body simulations of the dynamical evolution of star-forming regions by \cite{parker13} also support this scenario. 
In these simulations, initially sub-structured super-virial agglomerations will evolve to form a multi-clustered 
region, similar to NGC\,346, within 5\,Myr.

The simulations by \cite{parker13} predict that star-forming regions will dynamically evolve to form bound clusters or 
unbound associations, depending on their initial conditions, i.e., their virial status and fractal dimension. Based on the 
present status of the central cluster, we cannot know which of the above scenarios explains its origins. Specifically, it is 
unclear if the  cluster is still under formation (through merging or not), and if it is undergoing core collapse that evaporates 
its low-mass stars due to gas expulsion  \citep{baumgardtkroupa07}, or through rapid dynamical interactions \citep[][]{allison10}.
Considering that all simulations discussed here deal with length-scales and stellar numbers smaller than that of NGC\,346, 
modeling of richer stellar samples in larger areas is surely necessary for constraining the initial conditions that led to the 
formation of NGC\,346.


Most simulations predict central segregation of the most massive stars, but they measure it with different methods and explain it 
with different mechanisms. Mass segregation  may be primordial because massive stars are born in situ at the central 
region, or due to rapid equipartition, and so it is the product of dynamical evolution. Mass segregation is observed in NGC\,346
through the comparison of its isochronal age with the time-scale for energy equipartition, i.e., for mass segregation \citep[][p. 74, see also 
\citealt{kroupa04}]{spitzer87}. \cite{sabbi08} found that the latter is one order of a magnitude larger, impliying that the observed mass 
segregation is likely due to initial conditions, rather than dynamical evolution.

The present dynamical status of whole region of NGC\,346 can be naively\footnote{Due to its asymmetrical stellar clustering, the determination of 
a crossing time for the whole region may not make sense. It is, though, useful in understanding its dynamical status, especially in comparison to other objects.} determined in terms of its relaxation and crossing time-scales. The relaxation 
time, $T_{\rm relax} \approx \left( 0.1\,N/\ln{N} \right)\,T_{\rm cr}$, is the time for significant energy redistribution to occur in a cluster 
\citep[][p. 37]{bt87}, where $T_{\rm cr} \approx 2r_{\rm h}/\sigma$, is the crossing time of a typical star through the cluster which has 
a characteristic radius $r_{\rm h}$ and a velocity dispersion $\sigma \approx GM/r_{\rm h}$ \citep[see also][]{kroupa08}. 
$G = 0.0045\,{\rm pc}^3/\left({\rm M}{\solar}\,{\rm Myr}^2\right)$ is the gravitational constant, $N$ the total number of stars, and $M$ that total 
stellar mass of the cluster. Using a radius $r_{\rm h} \simeq$\,9\,pc and a total mass $M \simeq 3.9\,10^{5}$\,M{\solar}, \cite{sabbi08} 
derives $T_{\rm relax} \simeq$\,570\,Myr, far larger than the age of NGC\,346. This result, supported by the apparent substructure in the region
suggests that NGC\,346 is not dynamically relaxed, in agreement with the assumption that mass segregation in NGC\,346 is primordial. 

\cite{gpz11} use the ratio of stellar age over the crossing time, 
$\Pi = {\rm Age}/T_{\rm cr}$, to distinguish bound from unbound stellar systems, with $\Pi < 1$ for unbound, i.e.,  expanding, objects. Using their
own definition of the crossing time in terms of empirical cluster parameters (their Eq. 1), and the values for age, $r_{\rm h}$ and
$M$ from Sabbi et al., they derive $T_{\rm cr} = 6.4$\,Myr, and thus $\Pi \approx 0.5$, classifying NGC\,346 as unbound agglomerate.
However, assuming a velocity dispersion of $\sigma = 10$\,km\,s$^{-1}$, typical for star-forming complexes of the size of NGC\,346, 
Sabbi et al. derives $T_{\rm cr} = 1.8$\,Myr, which classifies NGC\,346 as a bound system. 

The difference in the results between the above studies is important, because they influence our understanding of the origin of the region and, 
thus, the nature of its bimodal stellar clustering.  If NGC\,346 is a gravitationally bound object, but not relaxed yet, as the findings of \cite{sabbi08} 
suggest, then probably the condensed stellar component is the {\em undergoing result of dynamical interactions} between smaller clusters. On 
the other hand if NGC\,346 expands as the result of \cite{gpz11} implies, then most probably the central cluster {\em was originally formed 
condensed}. Under these circumstances a clarification about the origin of the central cluster and of the bimodal clustering behavior of the young 
stars in NGC\,346 can only be achieved with the accurate determination of stellar dynamics, i.e., kinematics, of the region. 


By reversing the above analysis we make a prediction for the velocity dispersion that the system would have if it was dynamically 
stable or unstable. The condition for 
stability for NGC\,346 ($\Pi = 1$) requires $T_{\rm cr} = 3$\,Myr, which implies a velocity dispersion of $\sigma \simeq$\,6\,km\,s$^{-1}$ 
for $r_{\rm h} = 9$\,pc. (This value is in excellent agreement with the line-of-sight velocity dispersion of stars within a projected distance 
of 5\,pc from the centre of the young massive cluster R136 in the Tarantula nebula; see \citealt{brunet12}.) It should be noted that the 
half-number radius, i.e., the characteristic radius, in our young stellar sample is larger than that determined by Sabbi\,et\,al., 
and equal to $r_{\rm h} \simeq 16.5$\,pc.
Assuming this radius, the velocity dispersion at the stability limit raises to $\sigma \simeq$\,11\,km\,s$^{-1}$, still within typical values \citep[e.g.,][]{sana13}.
The predicted values of 6\,-\,11\,km\,s$^{-1}$ for the velocity dispersion of NGC\,346 provide a limit for characterizing the stability of the system as a whole. 
If  the velocity dispersion is larger that this limit, NGC\,346 is bound and collapsing (since $\Pi > 1$), while for values lower than this estimate, 
the system is unbound, and thus possibly under dissolution. In any case, different stellar components in the region are expected to demonstrate 
different velocity dispersions, depending on their own dynamical status.

In the above discussion we assume a single age of $\sim$\,3\,Myr  for the entire region, based on its PMS stars \citep{sabbi08}. 
However, single-epoch photometry of PMS stars is significantly affected by their physical characteristics (unresolved binarity, accretion, 
circumstellar extinction, variability, etc) that dislocate these stars from their theoretical positions on the color-magnitude diagram. 
This effect causes a color-luminosity spread of the PMS stars, which translates to an artificial age-spread, making the measurement 
of stellar ages (and masses) quite uncertain \citep[see review by][]{gouliermis12}. This issue, for {\sl Hubble} imaging of PMS stars
in the Magellanic Clouds, can only be accurately addressed through probabilistic determination of their physical parameters 
\citep[e.g.,][]{dario10}. Naturally, a more accurate determination of PMS ages in NGC\,346 and verification of an age-spread, which may even 
be positional dependent, would have noticeable implications to our understanding of star formation in the region.

\section{Concluding Remarks}\label{s:concl}


Concluding remarks, derived from our analysis, are summarized to the following: 

\begin{description}


\item 
Large star-forming complexes are excellent stellar `ecosystems' for the investigation of clustered star formation, because
the represent the typical 100-pc scale of giant molecular clouds.

\item 
The combined application of various cluster analysis tools is necessary for thoroughly characterizing young stellar clustering. 
In particular the autocorrelation function emerges as a  robust method, because, supported by the appropriate simulations, it 
is capable to distinguish different clustering styles.

\item 
We determine that the clustering of young stellar populations in the star-forming complex NGC\,346 has 
two distinct components; An extended self-similar, i.e., hierarchical, stellar distribution, and at least a centrally 
condensed young stellar cluster.

\item 
Dedicated simulations of combined stellar distributions show that the condensed stellar component of
fits to a spherical cluster, including 40\% of the stars, with a core radius between 
8\arcsec and 9\arcsec ($\sim$\,2.5\,pc) and a power-law surface density profile with slope between 2.20 and 2.35. The
remaining population is fractally distributed across the whole extent of the region with a fractal dimension $D_3$
between 2.20 and 2.36.

\item 
Our findings suggest that the present clustering behavior of young stars in NGC\,346 is the product of both star 
formation (turbulent-induced hierarchy), and early dynamical evolution (merging towards or dissolution of a 
centrally condensed cluster). Both processes seem to be still active after a time-scale of $\sim$\,3\,Myr, i.e., the 
assumed evolutionary age of the region.

\item 
The origin of this bimodal clustering behavior is not clearly understood. Considering that it influences the future evolution of the
region, we assess that if the complete system is under disruption then possibly the central cluster was formed condensed. 
If NGC\,346 contracts under its own gravity then this would imply that the central cluster may be the product of this contraction.
We determine that the velocity dispersion limit for stability of the region as a whole is $\sigma \gsim$\,6\,-\,11\,km\,s$^{-1}$. However, 
based on the heterogeneous clustering of the region, the {\sl local} velocity dispersion of individual sub-structures should demonstrate
variations.


\end{description}

Considering the excellent coincidence of the fractal dimension derived for the stellar clustering with that measured and 
theoretically predicted for the turbulent interstellar medium, a natural step forward for this study is to understand how 
the stellar assembling process relates to the structuring behavior of the natal interstellar medium. This topic will be 
addressed in a subsequent study, where we compare the interstellar gas structure to that presented here for the young 
stars in NGC\,346.

\section*{Acknowledgments}
We thank Richard Parker and Stefan Schmeja for their critical views and useful discussions that helped improving our interpretation. 
D.A.G. kindly acknowledges financial support by the German Research Foundation through grant GO\,1659/3-1. 
S.H. and R.S.K. acknowledge support from the Collaborative Research Center ``The Milky Way System'' (SFB\,881), 
particularly subproject B5, of the German Research Foundation. Based on observations made with the NASA/ESA {\sl Hubble 
Space Telescope}, obtained from the data archive at the Space Telescope Science Institute (STScI). STScI is operated by the 
Association of Universities for Research in Astronomy, Inc.\ under NASA contract NAS 5-26555. 



\appendix

\section{Library of Simulated Autocorrelation Functions}\label{s:appendix}


\subsection{The Requirement for Simulated ACFs}\label{s:acflim}

A cluster analysis method, developed by \cite{larson95} as a modification of the standard 
two-point angular correlation function \citep{gomez93}, is the correlation of the mean surface 
density of stellar companions (MSDC) against pair separation (in logarithmic scales). 
In general, the MSDC behaves in analogy to the ACF\footnote{An advantage of the 
ACF over the MSDC is that the former is normalized by the average density in the 
survey area. This allows the clear distinction between strongly and weakly clustered samples
(from the absolute ACF values), and the direct comparison of the shape of ACF for different 
ensembles.}, with a power-law index $\eta$ being identically associated to the fractal dimension 
$D_2$. Both methods, however, are subject to observational limitations in their interpretation. 

A break in the power-law of the MSDC of Galactic star-forming regions is suggested to be strongly 
influenced by the overall stellar surface density \citep{simon97, bate98}, rather than representing a 
characteristic scale, as was previously proposed. Moreover, a single power-law index 
may not even be related to the fractal dimension of the clustering, merely reflecting the large-scale density 
gradient in a centrally concentrated cluster \citep{bate98, klessenkroupa}. These limitations,
in addition to edge effects introduced by the unavoidably limited observed fields \citep[][see also 
Section\,\ref{s:edgeeffect}]{cw04}, require the simulation of distinctive stellar clusterings 
for the correct characterization of the ACF, and its use as a cluster analysis diagnostic.


In this Appendix we build a library of typical ACFs, based on simulations of 
centrally concentrated clusters and fractal stellar distributions. We 
describe these simulations and their ACFs in Appendices\,\ref{s:ccclusters} and \ref{s:fractaldistr}, where 
we also verify the behavior of the ACF for a random stellar distribution, i.e., for a sample of 
non-clustered stars (Appendices\,\ref{s:king}\, and\,\ref{s:effclusters}). 
These `field'  ACFs are consistent with a value of $\simeq$\,1, and remain 
unchanged with the separation range. 
The spatial coverage and stellar numbers of all considered artificial distributions are scaled to be 
identical to the field-of-view and stellar sample covered by our {\sl Hubble} ACS observations of NGC\,346.

\subsection{The Autocorrelation Function of Centrally Concentrated Stellar Distributions}\label{s:ccclusters}


We examine the autocorrelation of stars spatially related to each other 
within a spherically symmetric cluster. We compose artificial centrally-condensed 
stellar distributions that follow typical stellar surface density profiles, and we 
construct the corresponding ACFs in order to comprehend their behavior in
comparison to the observed ACF of NGC\,346. All simulations are performed 
for the same number of stars with that in our observed sample (5,150) confined 
within a field-of-view identical to that covered by the three ACS/WFC pointings mosaic.
We consider two types of radial stellar density profiles for the synthetic clusters:
Profiles following the empirical law by \cite{king62},  
and those represented by the model of \cite{eff87}. The latter, representing clusters 
that are not tidally truncated, are 
best-suited for extended clusters surrounded by well populated stellar fields, as is usually 
the case in large star forming regions like NGC\,346. Therefore we base our analysis on 
these clusters, the simulations of which are presented in Appendix\,\ref{s:effclusters}. 
Below, we 
also present the King star cluster profiles for reasons of completion in 
our study, and for a reference for future studies on the ACF of star clusters.

\subsubsection{King-profile Clusters}\label{s:king}

We further investigate the ACF of more realistic centrally concentrated 
stellar clusters in dynamical equipartition. Artificial spherical clusters were simulated 
to follow stellar surface density profiles defined by King's semi-empirical model 
\citep[e.g.,][]{king62}. The functional form of these profiles is: 
\begin{eqnarray}\label{eq:king} 
f(r) & \propto & \Bigl(\frac{1}{[1+(r/r_{\rm c})^2]^\frac{1}{2}}-\frac{1}
{[1+(r_{\rm t}/r_{\rm c})^2]^\frac{1}{2}}\Bigr)^{2}, 
\end{eqnarray} where $f$ is the stellar surface density, and $r$ is radial distance.
The clusters are constructed with various values in their concentration 
parameter $c$, defined as $c=\log{r_{\rm t}/r_{\rm c}}$, where $r_{\rm t}$ and 
$r_{\rm c}$ are the tidal and core radius of the cluster. 

\begin{figure}
\centering
\includegraphics[width=\columnwidth]{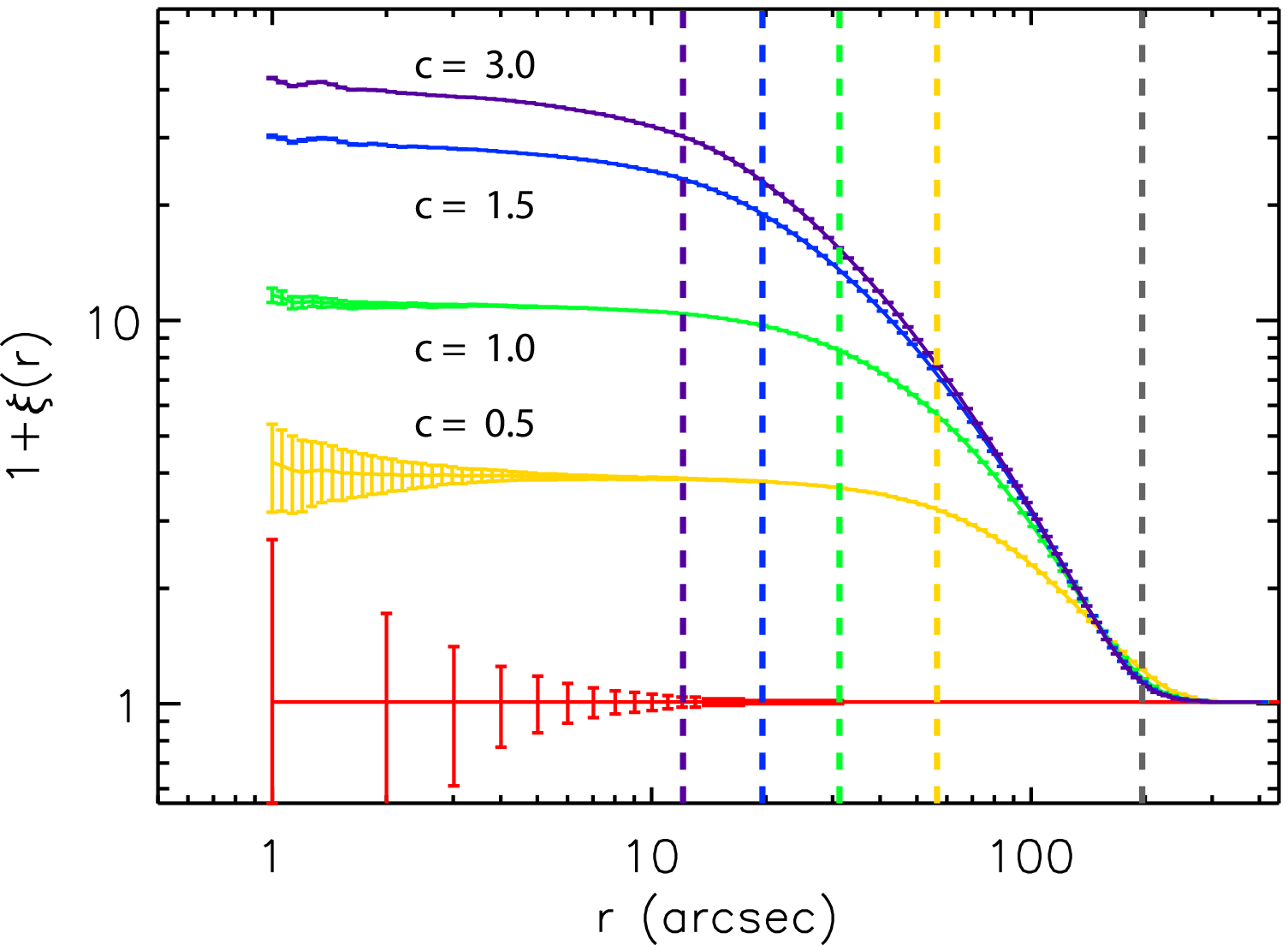} 
\caption{ACFs for a sample of four centrally 
concentrated clusters that follow King-type stellar density 
profiles with different concentration parameters, $c$, as indicated in the plot. Vertical 
color dotted lines correspond to the $r_{\rm h}$ of the clusters, and the black dashed line 
corresponds to the $r_{\rm t}$ of the clusters, which is chosen to be identical and equal to 
55\,pc. The ACF of a random stellar distribution is plotted for reference in red.}
\label{fig:acfmodel}
\end{figure}

The ACFs for a sample of such clusters are shown in Figure\,\ref{fig:acfmodel} 
along with that for the random field (plotted in red). Clusters are constructed with concentration parameters 
equal to 0.5 (plotted in orange), 1.0 (light green), 1.5 (green) and 3.0 (blue) respectively.
The latter represents the observed extreme in the parameters space of
known Galactic globular clusters \citep[Pal\,12, $c=2.98$;][2010 Edition\footnote{
\href{http://physwww.physics.mcmaster.ca/~harris/mwgc.dat}{http://physwww.physics.mcmaster.ca/$\sim$harris/mwgc.dat}}]{harris96}.
From these plots it is seen that while indeed for all clusters the autocorrelation
function is $1+\xi(r) > 1$, it does not again remain constant through the whole range 
of stellar separations $r$, but it drops at larger separations towards the outskirts of the 
clusters converging to the random field value of unity. The ACFs reach this value 
at the radii where essentially the clusters meet the surrounding field, comparable 
to their tidal radii. For reasons of simplicity, in the examples shown 
in Figure\,\ref{fig:acfmodel} all  clusters are constructed to have the same 
$r_{\rm t} = 55$\,pc ($\simeq$\,196\arcsec). 

\begin{figure}
\centering
\includegraphics[width=\columnwidth]{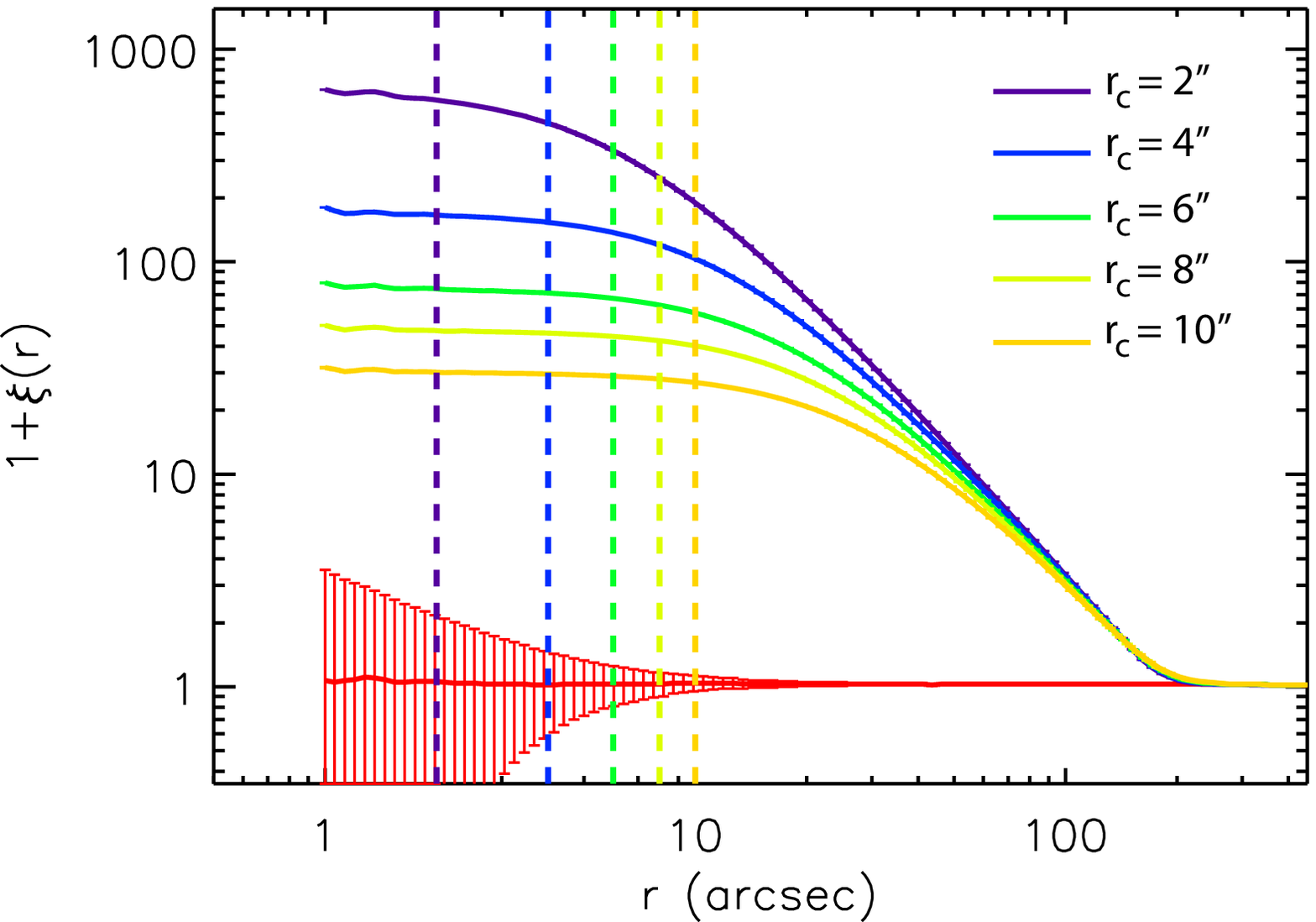} 
\caption{ACFs for a sample of five centrally concentrated clusters that follow EFF-type 
stellar density profiles. All clusters in the plot have the same number of stars, $\sim5,000$, 
and a fixed outer profile slope, $\gamma=3$, typical for young clusters in the Magellanic 
Clouds, while their core radii, $r_{\rm c}$, vary. They are selected to have 
typical values of 2\arcsec, 4\arcsec, 6\arcsec, 8\arcsec, and 10\arcsec, indicated with vertical 
dashed lines in magenta, blue, green, light-green and orange respectively. The
ACF of a simulated random (unclustered) field of the same number of stars is shown in red
for reference.}
\label{f:effacs1}
\end{figure}

\begin{figure}
\centering
\includegraphics[width=\columnwidth]{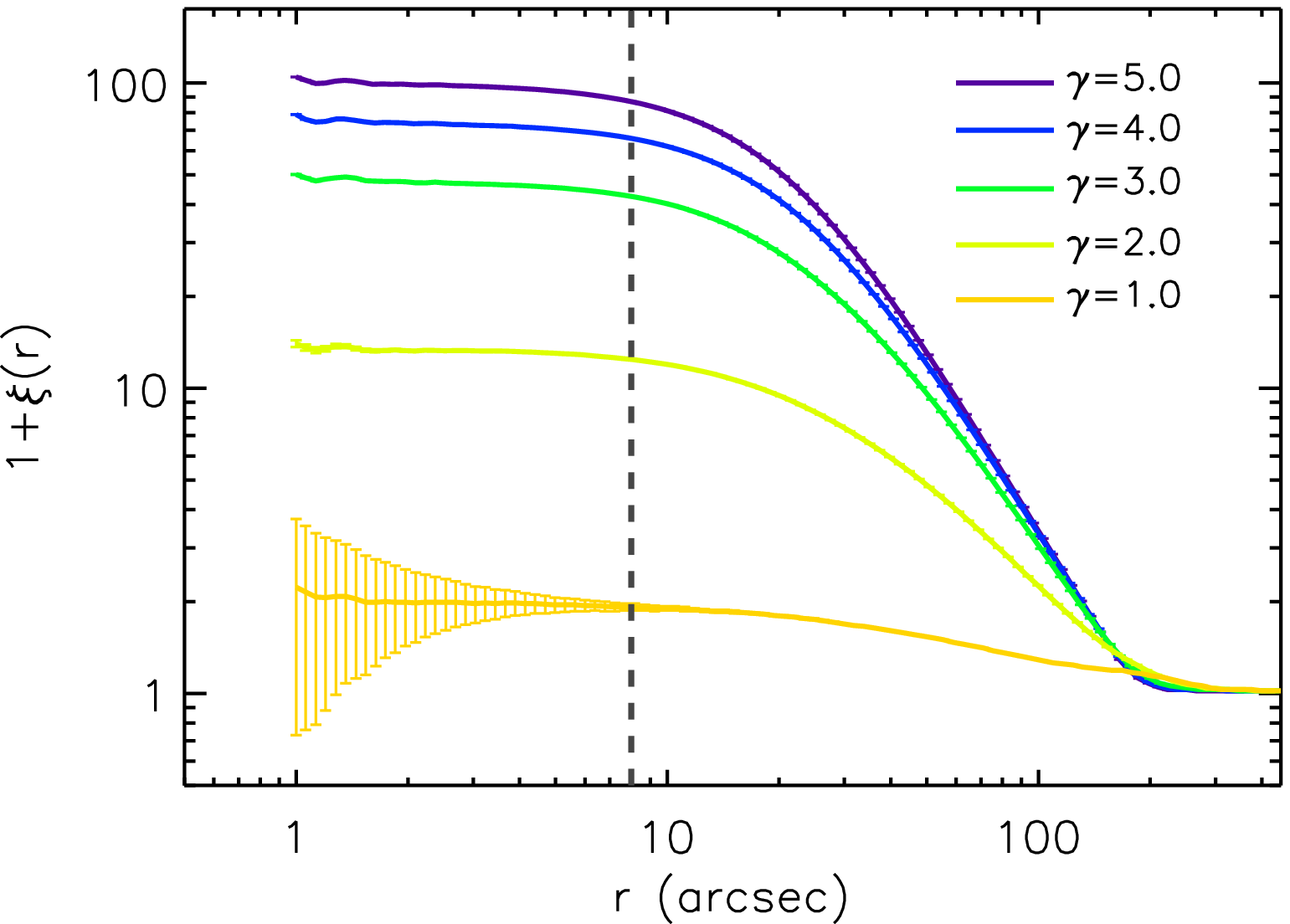} 
\caption{ACFs for a sample of five centrally concentrated clusters that follow EFF-type 
stellar density profiles. All clusters have the same number of stars, $\sim5,000$, 
and a typical fixed core radius, $r_{\rm c}=8\arcsec$, indicated in the plot with the 
vertical grey dashed line. Their outer density profile indexes, $\gamma$, are selected 
to have values typical for young Magellanic Clouds clusters, of 1, 2, 3, 4, and 5. The case
of $\gamma=1$ is an unobserved extreme used here for demonstrating the flatness of
its ACF.}
\label{f:effacs2}
\end{figure}

There are two derivatives from these simulations: (1) The maximum values of $1+\xi$ 
depend on the degree of concentration, i.e. the concentration parameter, of each cluster; 
More centrally concentrated clusters have higher ACF values, and they are thus `more clustered'. 
This trend, which is more 
obvious at small separations, becomes less important for clusters with $c \gsim 1.5$. (2) 
The shape of the ACF drops at larger separations (up to the field value) also depends 
on the concentration parameter. It naturally depends also on the limiting radius of the cluster, 
i.e., its  tidal radius; Clusters with larger $r_{\rm t}$ and smaller $c$ have smoother drop in their 
ACFs. In the examples shown in Figure\,\ref{fig:acfmodel} all clusters have the same $r_{\rm t} $, and thus the 
steepness of the drop of their ACF depends only on their concentration parameter. 

\begin{figure*}
\centering
\includegraphics[width=0.975\textwidth]{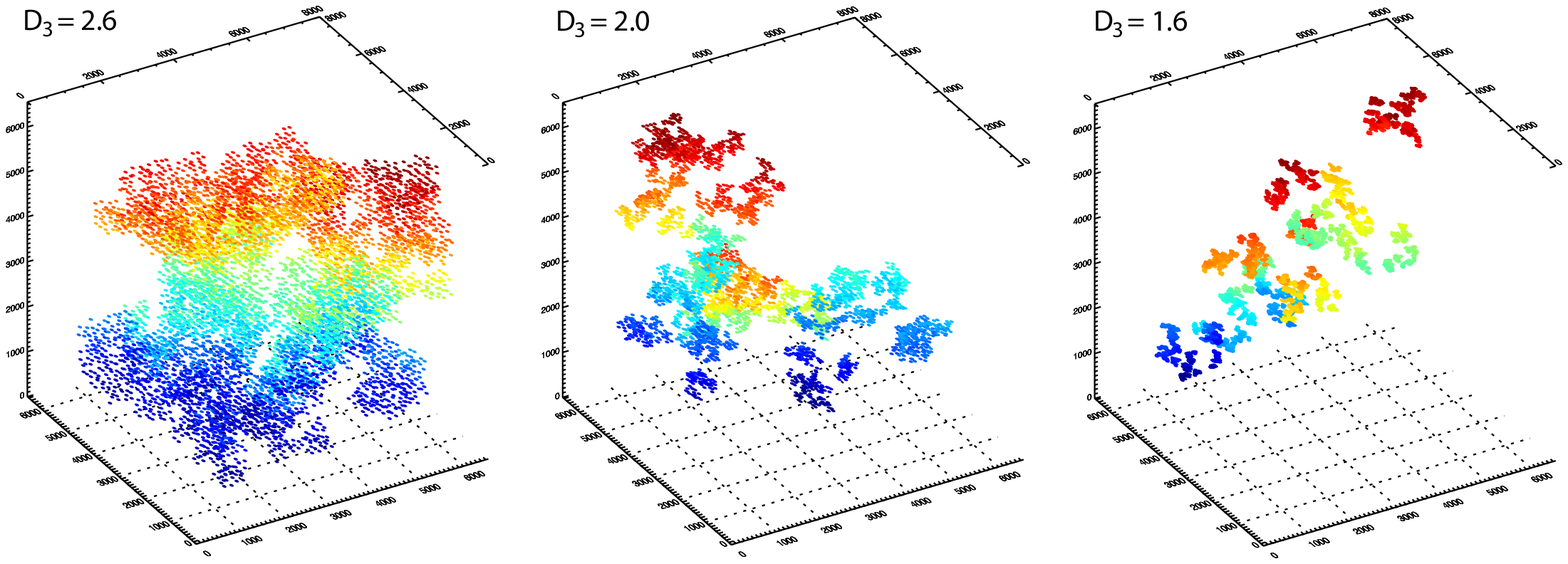} 
\caption{Sample of our simulations of three  self-similar stellar distributions. 
The distributions of few thousand stars with fractal dimensions $D_{3} =$\,1.6,\,2.0\,and\,2.6
are shown. Note the trend of stellar clustering to higher clumpiness as $D_3$ becomes
smaller, i.e., self-similarity becomes stronger. On the other hand, the fractal dimension of $D_3=2.6$ is very close
to the geometrical dimension, and thus the corresponding distribution is very close to a random. 
Points are colored according to their $z$-axis positions.}
\label{f:frctsmpl}
\end{figure*}

\subsubsection{EFF-Profile Clusters}\label{s:effclusters}

The outskirts 
of young stellar clusters, located in star-forming regions of the Magellanic Clouds,  
demonstrate extended outer envelopes, which cannot be represented by King's semi-empirical 
model \citep[e.g.,][]{king62}, designed for tidally truncated globular clusters. 
\cite{eff87} developed an empirical model more suitable to describe the stellar surface density 
profile of such clusters (from hereon the EFF model). We base our simulations of 
centrally-concentrated clusters on this model.

The surface stellar density of the cluster according to the EFF model is described as: 
\begin{eqnarray}\label{eq:eff}
f(r) & = & f_{0} {\left[ 1 + {\left( \frac{r}{\alpha}
\right)}^{2} \right]}^{-\gamma/2}+f_{\rm{field}},
\end{eqnarray} 
where $f_{0}$ is the central stellar surface density, $\alpha$ is a measure of the core 
radius and $\gamma$ is the power-law slope which describes the decrease of 
surface density of the cluster at large radii; $f(r)\propto r^{-\gamma/2}$ for $r\gg a$.  
The uniform background density is given by $f_{\rm field}$. For comparison, 
a King profile (Eq.\,\ref{eq:king}) for $r_{\rm t} \gg r_{\rm c}$ would be described as:
\begin{eqnarray}\label{eq:kingeff}
f(r) & = & f_{0} {\left[ 1 + {\left( \frac{r}{r_{\rm c}}
\right)}^{2} \right]}^{-1}+f_{\rm{field}}.
\end{eqnarray} 
The basic parameters of the EFF model, $\alpha$, $\gamma$ and $f_{\rm field}$, are measured 
from the fitting of observed profiles with the function of Eq.\,(\ref{eq:eff}). 

In order to determine the ACF of clusters following EFF profiles, and compare it to that for 
NGC\,346, we simulate clusters that contain the same number of stars as the observed field
of NGC\,346 placed in a comparable field-of-view. Apart from the total number of stars of the cluster,
we construct our EFF centrally concentrated 
clusters by providing as input parameters its core radius, $r_{\rm c}$, and the outer surface 
density index, $\gamma$. For reasons of simplicity we ignore the constant contribution 
of the field, assuming $f_{\rm field} = 0$.  

According to the EFF model, the relation of 
$\alpha$ to the core radius is given from Eq.\,(\ref{eq:eff}) assuming no contribution from 
the field: 
\begin{eqnarray}
r_{\rm c} & = & \alpha(2^{2/\gamma}-1)^{1/2} \label{eq-rc}. 
\end{eqnarray}
Our simulated clusters have $r_{\rm c}$ and $\gamma$ values comparable to typical 
values for young clusters in the Magellanic Clouds \citep{mg03lmc, mg03smc}. In Figure\,\ref{f:effacs1}
we show the ACFs of five EFF clusters with the same typical value of $\gamma = 3$ and different 
core radii. Figure\,\ref{f:effacs2} demonstrates the constructed ACFs for such clusters keeping the 
core radius constant at the typical value of $r_{\rm c} = 8\arcsec$ while varying the profile
index $\gamma$ within the range of observed values.

Both Figures\,\ref{f:effacs1} and\,\ref{f:effacs2} demonstrate the dependance of the 
shape of the ACF of centrally condensed clusters on the assumed structural parameters
of the clusters. In Figure\,\ref{f:effacs1} can be seen 
that, for clusters with the same number of stars, larger $r_{\rm c}$ produced 
more shallow ACFs in particular at smaller separations. The reason for this behavior
is that stars would be confined in a narrow area (with short separations) if the core 
were small. A larger core for the cluster forces them to be distributed over a wider area 
with higher separations among them, flattening the ACF at small separations. These
clusters will appear in the ACF plot `less clustered' than those which are more centrally
concentrated, i.e., in smaller cores.  
On the other hand, as shown in Figure\,\ref{f:effacs2}, if the core remains unchanged, the 
surface density profile slopes change accordingly the decrease of the ACF at large 
separations. Steeper density profiles lead to ACFs, which are steeper at large separations.
The slope of the ACF at smaller separations remains unchanged, while its absolute values 
are smaller (less clustered) for more shallow stellar surface density profiles at the outskirts
of the clusters.

Both figures with the ACFs of EFF-profile clusters show a change in their separations dependence, 
i.e., a `breaking' of the ACF slope, at different scales. In Figure\,\ref{f:effacs1}, the use of clusters with the 
same $\gamma$ demonstrates that the radial distance where the separation dependence of ACF 
changes, i.e., where the power-law `breaks', depends on how well concentrated the clusters are, i.e.,
on their $r_{\rm c}$. Therefore, the separation, where the break occurs may be related to a radial 
scale, which is characteristic for its stability\footnote{We explored this dependance for clusters following a 
King profile in Appendix\,\ref{s:king}.}. 

\subsection{The Autocorrelation Function of Self-Similar Distributions}\label{s:fractaldistr}

\begin{figure}
\centering
\includegraphics[width=\columnwidth]{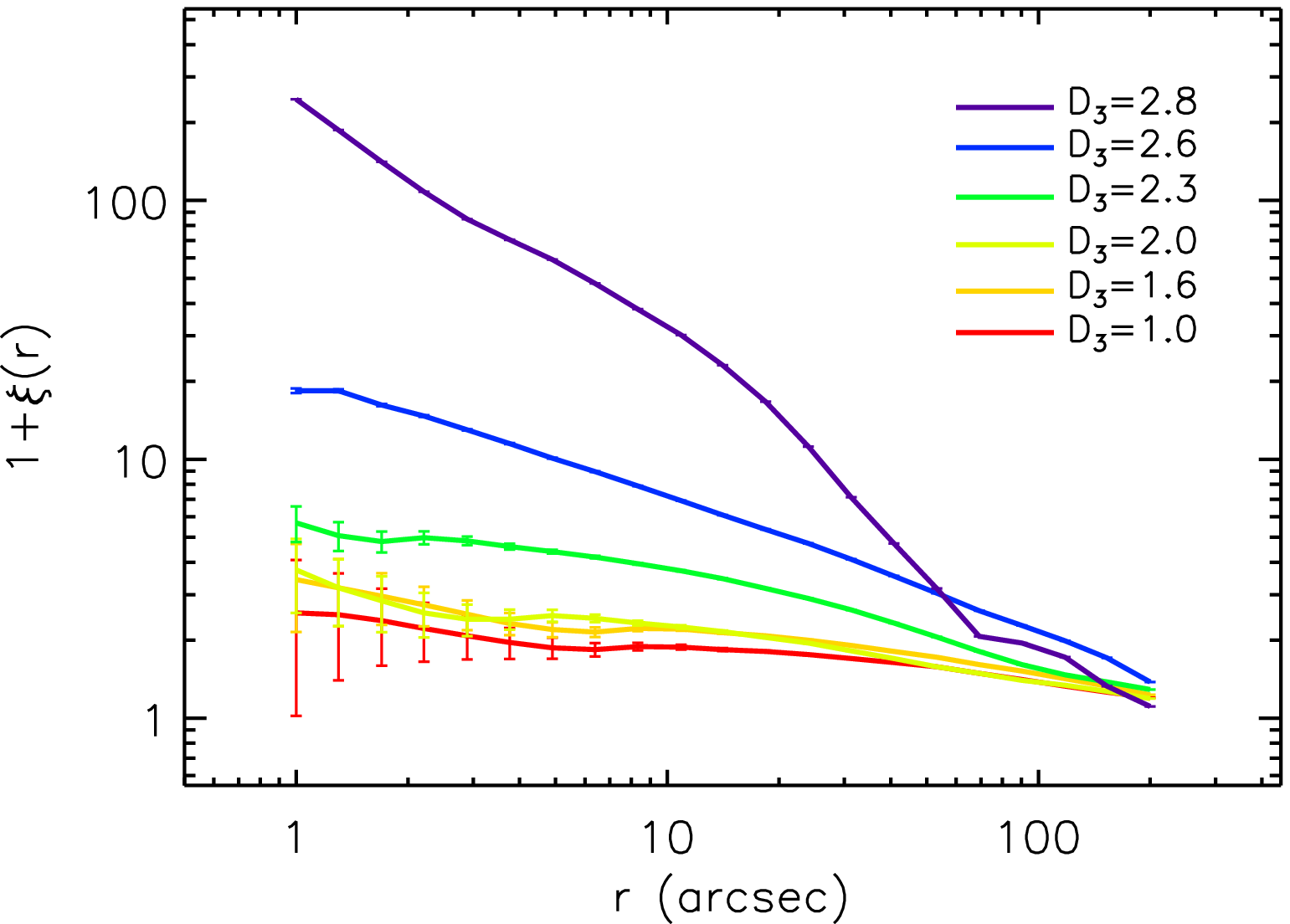} 
\caption{ACFs of six different self-similar stellar distributions, having 
fractal dimensions of $D_{3} =$\,1,\,1.6,\,2,\,2.3,\,2.6\,and 2.8. A sample of one 
distribution out of several constructed for each $D_{3}$ is shown here. These 
plots show that the ACFs of fractal stellar concentrations should monotonically 
increase for smaller stellar separations, unlike what we observe in NGC\,346.}
\label{fig:acfsfractal}
\end{figure}

In this section we consider the ACF of self-similar, i.e., fractal stellar
distributions. We construct three-dimensional artificial fractal distributions 
with the application of a reverse box-counting algorithm by first defining a 
cube of side-length 1. Next, this cube is divided into ${\cal N}_{\rm div}^3$ 
equal sub-cubes, of side-length $1/{\cal N}_{\rm div}$, which is the first 
generation of `children'. ${\cal N}_{\rm ran}$ of these sub-cubes are then 
randomly selected to become parents themselves and further divided into
${\cal N}_{\rm div}^3$ child sub-cubes, and the process is repeated 
recursively, terminating at the desired level of recursion. At this stage 
each of the smallest (final generation) sub-cubes has a star placed in it. 
Finally, we normalize the three-dimensional coordinates of these
stars to a total side-length of the order of our observed 
field-of-view\footnote{We assume an average side-length of 6656 pixels, 
corresponding to $\simeq$\,5.55\arcmin, or $\simeq$\,94\,pc.}.
The counting-box fractal dimension depends on both ${\cal N}_{\rm div}$
and ${\cal N}_{\rm ran}$, and is defined as
\begin{eqnarray}\label{eq:cboxfract} 
D \equiv \frac{\ln{({\cal N}_{\rm ran})}}{\ln{({\cal N}_{\rm div})}}.  
\end{eqnarray}
Normally we use ${\cal N}_{\rm div} = 2$, in which case 
there are 8 sub-cubes (children) in every generation. 
The probability that a child cube will further be divided and become a parent 
is ${\cal N}_{\rm div}^{(D-3)}$. For lower $D$, this probability is lower and the 
distribution becomes more `porous'. This technique was originally implemented by various authors 
\citep[see, e.g.,][]{bate98, cw04}.

The produced fractal distributions contain 
specific stellar numbers by culling randomly the stars that populate the final 
samples. The number of iterations, i.e., of 
produced generations, depends on the total number of stars to populate 
the distribution, $N_{\star}$ and ${\cal N}_{\rm ran}$ as $N_{\rm gen} = 
\log{(N_{\star})}/\log{({\cal N}_{\rm ran})}$. Normally we require 10,000 stars 
to populate our fractal distributions, and thus we produce about 13 generations
in each simulation. To avoid an obviously regular structure, we add a bit noise to 
the positions of the stars, by repositioning every star by a random infinitely small 
fraction of the sub-cube size. The examples of three such fractal distributions are 
shown in Figure\,\ref{f:frctsmpl}.

The ACFs of indicative self-similar distributions of approximately 5,000 stars with fractal dimensions 
$D_3$ between 1.0 and 2.8, are shown in Figure\,\ref{fig:acfsfractal}.
These ACFs show a monotonous decrease with stellar separation, and no 
indication of a change of slope at a specific scale, unlike what we find in NGC\,346.
We examined the effect of different viewing angles of the distributions to their 
two-dimensional projections and the subsequent effect to the produced ACF, and
we found that -- at least for these stellar numbers of $\sim$\,5,000 -- the derived 
ACFs show no difference at all, with the corresponding indexes remaining essentially unaffected.
It should be noted that the pure 3D fractals generated by the box counting method,
after projected on the 2D plane, produce a ``wiggle'' in their ACF, which nevertheless does not 
affect their monotonous trend. The scale where this wiggle appears seems to 
depend on the fractal dimension, occurring for low fractal dimensions at larger scales than
for higher fractal dimensions. In the case of the best representative simulations ($D_3$ $\sim$ 
2.3) this wiggle occurs at separations \lsim\,2\arcsec.


\section{Relation between the Fractal Dimensions $D_3$ and $D_2$ through the ACF Index $\eta$}\label{d3toeta}

The application of the ACF provides a measurement of the two-dimensional fractal dimension 
$D_2$ of the considered ensemble through its index $\eta$ (see, e.g., Section\,\ref{s:acf}). However,
we construct our simulated fractal distributions in a volume, providing as basic input parameter the 
three-dimensional fractal dimension $D_3$, and there is no direct relation between $D_3$
and $D_2$. A simple conversion $D_3=D_2+1$ is usually cited, which, however, applies only 
if the perimeter-area dimension of a projected 3D structure is the same as the perimeter-area 
dimension of a slice \citep[][]{elmegreenscalo04}. We utilize our simulations of self-similar stellar distributions to 
provide a more general conversion between $D_3$ and $\eta$, and consequently between $D_3$ and $D_2$. 
This empirical conversion can be very useful for comparing results from methods 
that measure $D_2$ to those from methods that provide measurements for $D_3$  \citep[see, e.g.,][for 
a discussion on the available methods]{federrath09}, with no need for assuming the simple relation 
$D_3=D_2+1$.

\begin{figure}
\centering
\includegraphics[width=\columnwidth]{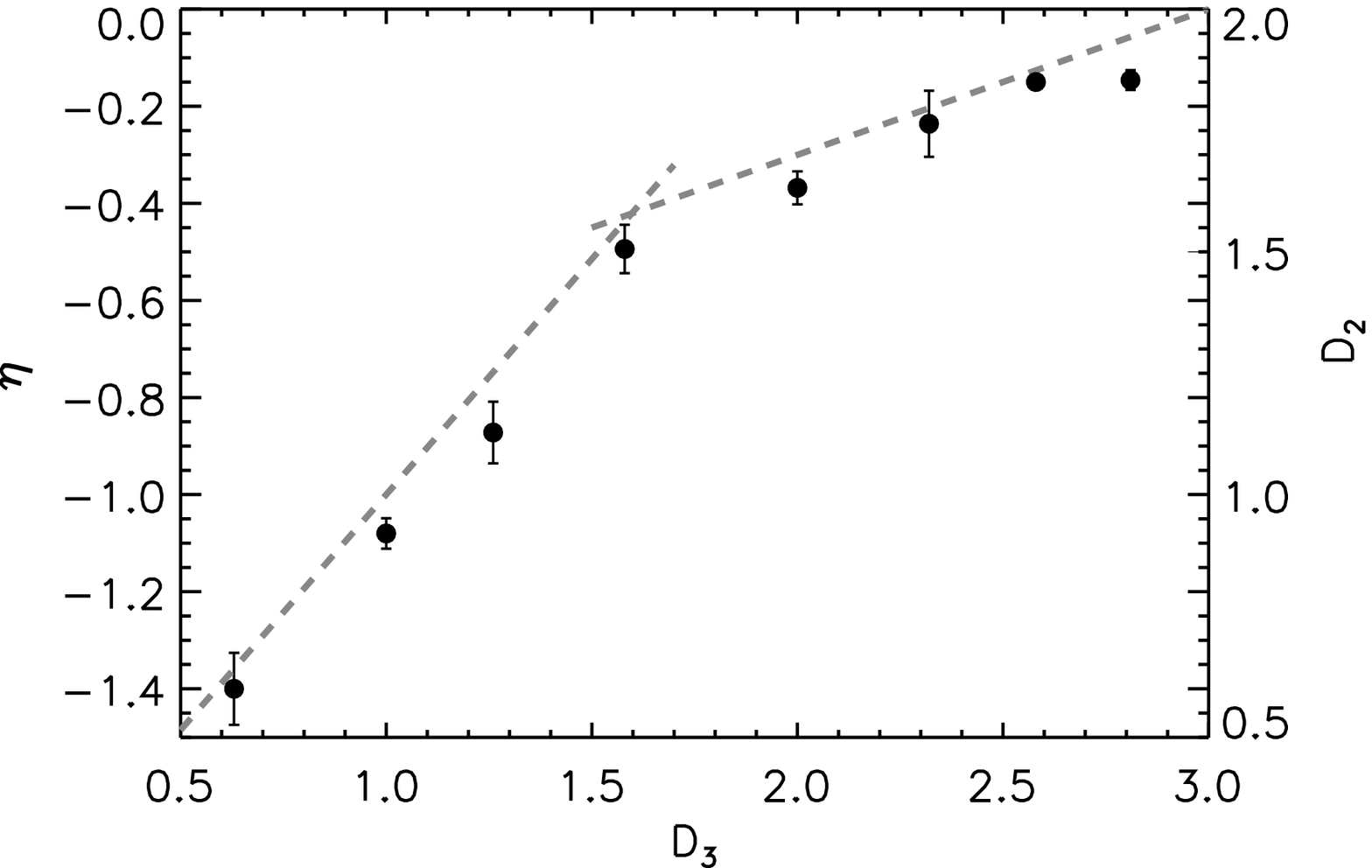}
\caption{Calibration relation between the three-dimensional fractal dimension 
$D_3$, the ACF index $\eta$, and the corresponding two-dimensional fractal dimension $D_2$,
derived from our simulated self-similar stellar distributions.}
\label{f:d3eatd2}
\end{figure}

Since in our box-counting simulations we have to use integer numbers for ${\cal N}_{\rm ran}$ 
and ${\cal N}_{\rm div}$, there is a limited number of input values for $D_3$ that can be applied.
We performed simulations assuming two numbers for ${\cal N}_{\rm div}$ (2 and 3) and established 
a set of 8 self-similar distributions with different $D_3$, which span the complete realistic range of 
values between $D_3 =$\,0.5 and \lsim\,3. Then we constructed their ACFs and determined the corresponding 
indexes $\eta$, as well as the corresponding two-dimensional fractal dimensions $D_2$ from the relation
$D_2 = 2+ \eta$. The derived empirical calibration is shown in Figure\,\ref{f:d3eatd2}.  From this figure  
one can see that $D_2$ almost equals $D_3$ only for $D_3$\,\lsim\,1.6; this one-to-one relation is represented 
by the left-hand dashed grey line. For larger values of $D_3$, the derived $D_2$ converges 
toward its maximum possible value of 2 while $D_3$ approaches its own maximum of 3, having another linear 
dependence to each other, demonstrated by the right-hand dashed grey line. A simple functional form that 
can thus represent this relation is:
\begin{eqnarray}
D_{2} \simeq \left\{
          \begin{array}{r@{\,\,}l}
          D_{3} & ,\,{\rm for}\, D_3 \le 1.6\\
          & \\
          1.1 + 0.3\,D_{3} & ,\,{\rm for}\, D_3 \ge 1.6\\
          \end{array}\right.
\label{eq:funcform}
\end{eqnarray}


It is worth noting that for typical
$D_3$ values derived for turbulent-induced hierarchy, measured in interstellar gas ($D_3 \sim 2.3$), the 
corresponding $D_2$ value according to our conversion is $D_2 \sim 1.8$. The errors shown in  Figure\,\ref{f:d3eatd2}
are the standard deviations of the derived values for $\eta$ after performing few realizations in the construction of the 
synthetic fractal distributions for each $D_3$ value.

\end{document}